\documentstyle[12pt,epsf]{article}

\begin{document}

\newcommand{\sheptitle}
{String Gauge Unification and Infra-red Fixed Points
in MSSM+X Models}

\newcommand{\shepauthor}
{B. C. Allanach$^1$ and
S. F.
King$^2$ \\ 
\vspace{\baselineskip}
{\small 1. Rutherford Appleton Laboratory, Chilton, Didcot, OX11 0QX, U.K.} \\
{\small 2. Physics Department, University of
Southampton, Southampton, SO17 1BJ, U.K.}}

\newcommand{\shepabstract}
{In order to achieve gauge unification at the string scale 
$M_X\sim 5\times 10^{17}$ GeV in the 
minimal supersymmetric standard model (MSSM)
it is necessary to add extra gauge non-singlet representations
at an intermediate scale $M_I<M_X$, leading to a class of
models which we refer to as MSSM+X models.
We perform a detailed analysis of a large class of MSSM+X
models and find that the number of 
(3,1) representations added must be
greater than the total of the number of (3,2) and (1,2)
representations.
Predictions of $M_I$, $M_X$ and $\alpha(M_X)$ are obtained for models
with up to 5 extra vector representations than the MSSM\@. 
Upper bounds on the
U(1) string gauge normalisation $k_1$ and the sum of the squares of
the 
hypercharge assignments of the extra matter are also obtained for the 
models. We also study the infra-red fixed 
point behaviour of the top quark Yukawa coupling
in a large class of MSSM+X models and
find that the low energy MSSM quasi-fixed point
prediction of the top quark mass
is more likely to be realised in these theories
than in the MSSM\@. In other words the top quark
tends to be heavier in MSSM+X models than in the MSSM\@.
The implementation of a
U(1)$_X$ family symmetry into MSSM+X models to account for the
Standard Model
fermion masses is discussed and a
particular viable model is presented.}

\begin{titlepage}
\begin{flushright}
RAL 96-004 \\ SHEP 96-04 \\ hep-ph/9601391 \\
\end{flushright}
\vspace{.4in}
\begin{center}
{\large{\bf \sheptitle}}
\bigskip \\ \shepauthor \\ \mbox{} \\
\vspace{.5in}
{\bf Abstract} \bigskip \end{center} \setcounter{page}{0}
\shepabstract
\end{titlepage}

\section{Introduction}
The unification of the gauge couplings in supersymmetric grand unified
theories (SUSY GUTs) at a scale $M_{GUT}\sim 10^{16}$ GeV \cite{GUTun}
is often regarded as a triumph of the MSSM\@. 
Proponents of SUSY GUTs emphasise that such unification leads
to a prediction of $\sin^2 \theta_W$ at the 1\% level, and
the fact that the strong coupling $\alpha_s(M_Z)$
tends to come out on the large side is accounted for 
by threshold effects at the GUT scale which could in principle
lower $\alpha_s$ to any desired value in the experimental range.
However there are well known
potential threats to SUSY GUTs arising from experimental 
proton decay constraints on the one hand
and theoretical doublet-triplet splitting naturalness problems 
on the other. These two potential threats
can both be kept at bay at the expense
of adding several large Higgs representations
with carefully chosen couplings. A further challenge to SUSY GUTs
is the question of Yukawa matrices, which again requires large Higgs
representations. All these questions must and can be addressed
simultaneously, and there do exist realistic models in the
literature \cite{HR}.

With the advent of string theory there is a different possibility
for unification: string gauge unification in which the gauge
couplings are related to each other at the string scale $M_X$
\cite{gaugeun}.
String theories give the relation \cite{kap}
\begin{equation}
M_X = 5.3 \times 10^{17} g_X \mbox{~GeV},
\label{stringscale}
\end{equation}
which is independent of the Kac-Moody level,
where $g_X$ is the unified gauge coupling at the string scale $M_X$.
In this framework it is not necessary to unify the couplings
into a SUSY GUT group, although it is possible to 
envisage a scenario in which a SUSY GUT can be ``derived'' from the
string~\cite{su5xsu5,level2}.
Such string derivations must yield the desired
large Higgs representations with the precise couplings required 
to avoid proton decay, obtain doublet-triplet splitting,
and yield realistic Yukawa matrices.
This is a technical feat which has not yet fully been
accomplished, although there has been some recent
progress in this direction \cite{level2}. The existence of
adjoint Higgs representations requires the use of
Kac-Moody level 2 Virasaro algebras or higher,
and while it is not impossible that
Nature uses these higher levels, the
simpler string models are based on level 1 algebras. These are the
so-called string inspired models in which simple GUTs such
as $SU(5)$, $SO(10)$ and so on are abandoned, and instead
the gauge couplings are unified at the string scale.

The simplest possible string-motivated model is 
clearly the minimal supersymmetric standard model (MSSM).
In this picture the MSSM (and nothing else) persists
right up to the string scale $M_X$. Naively such theories
do not appear to be viable since we know that the
gauge couplings cross at $\sim 10^{16}$ GeV,
and will have significantly diverged at the string scale
$\sim 5\times 10^{17}$ GeV. However the situation is in fact not 
so clear cut since the $U(1)_Y$ hypercharge gauge coupling
has an undetermined normalisation factor $k_1\geq 1$ 
(where for example $k_1=5/3$ is the usual GUT normalisation)
which may be set to be a phenomenologically desired value
\cite{ibanez} by the choice of a particular string model. However the
simplest string theories
(e.g.\ heterotic string with any standard compactification)
predict equal gauge couplings for the other two observable sector
gauge groups
$g_2=g_3$ at the string scale $M_X$, which would require
a rather large correction in order to
account for $\alpha_s(m_Z)$~\cite{kap,thresholds}. In fact,
a recent analysis \cite{Dinesetal}
concludes that string threshold effects are insufficient by themselves
to resolve the experimental discrepancy. The analysis also concludes
that light SUSY thresholds and two loop corrections
cannot resolve the problem, even when acting together.
In order to allow the gauge couplings to unify at the string scale
it has been suggested \cite{ellisetc} that additional exotic matter
should
be added to the MSSM at some intermediate scale or scales $M_I<M_X$,
leading to a class of models which we shall refer to
as MSSM+X models.

The purpose of the present paper is twofold.
Firstly we shall perform a general unification analysis of a 
particular class of MSSM+X model. Then we shall
study the infra-red fixed point properties of such models,
focusing in particular on the top quark mass prediction.
A detailed unification analysis of 
MSSM+X models has also been performed by Martin and Ramond
(MR) \cite{MR}. MR considered the case of one or multiple intermediate
thresholds, where the intermediate matter was contained in incomplete
vector-like representations of $E_6$, either from chiral
or vector supermultiplets. Gauge extensions at the intermediate scale
were also considered \cite{MR}. 
The present unification analysis differs from the MR analysis 
in a number of ways as follows. Unlike MR,
we shall consider arbitrary numbers of chiral superfields
in low-dimensional vector representations, without any reference to
an underlying $E_6$ model. Furthermore, unlike 
ref.\cite{MR}, we shall not assume a GUT-type normalisation of the
hypercharge generator but instead allow the possibility of different
normalisations. Thus our analysis of string gauge unification
is complementary to that of MR\@. Turning to our
infra-red fixed point analysis of MSSM+X models, 
which was not considered at all by MR,
we shall focus attention on the infra-red fixed point
and quasi-fixed point of the top quark Yukawa coupling
within the above class of MSSM+X models 
using similar techniques to those proposed for the
MSSM and GUTs \cite{PR,H,RLfixed}. 
The main result is that the top quark mass tends to be
heavier than in the MSSM, and closer to its quasi-fixed point
in these models. Finally we speculate on the 
origin of Yukawa matrices with texture zeroes and small non-zero
couplings within the MSSM+X framework, using the idea of a $U(1)$
gauged flavour symmetry and multiple Higgs doublets at the
intermediate 
scale, similar to the proposal of Ibanez and Ross \cite{IR}.

The layout of the rest of the paper is as follows.
In section 2 we shall discuss string gauge unification 
of MSSM+X models, while in section 3 we shall consider the
infra-red stable fixed point of the top quark Yukawa
coupling in these models, and compare our results to 
those of the MSSM\@. In section 4 we shall briefly discuss the
possibilities
of obtaining realistic Yukawa matrices in this framework,
identifying the intermediate matter with multiple Higgs doublets
which, together with a $U(1)$ gauged flavour symmetry, may be used
to generate realistic textures.
Section 5 concludes the paper.

\section{String Gauge Unification in MSSM+X Models}
In this section we shall define the class of MSSM+X models
under consideration and discuss our calculational procedures
and the resulting predictions arising
from string gauge unification in these models.
Although there is inevitably some overlap in this
section with ref.\cite{MR}, we include our analysis in
detail since as discussed above, 
our starting point is somewhat different. 
Furthermore the string scale,
the intermediate scale and the 
string coupling will all be iteratively determined in our approach,
leading to predictions for these quantities
within particular models. Finally these results will be
necessary for the discussion of the infra-red fixed points
in the next section.

We shall impose string gauge unification subject to the
following restrictions and assumptions:
it is assumed that the gauge symmetry of the vacuum between the SUSY
breaking scale
$M_{SUSY}$ and the string scale is
SU(3)$\otimes$SU(2)$_L\otimes$U(1)$_Y$ and 
that the string theory is one
of Kac-Moody level 1. The last assumption allows us to
restrict the gauge representations since Kac-Moody level
1 strings only allow fundamental representations of the gauge group
for the matter representations. Thus, the only possible extra matter
representations 
we may add to the theory below $M_X$ are (3,1), (1,2) and
(3,2)\footnote{And other representations with either the 2 and
or the 3 conjugated. This point makes no difference to our analysis.}
representations in (SU(3),SU(2)$_L$) space. The constraint of
anomaly cancellation leads 
us to only add each of these representations to the MSSM
in complete vector representations.
We also assume for predictivity that the extra matter lies
approximately at
one mass scale $\sim M_I$. While this strong assumption is exact for 1
extra vector representation, it may be deemed increasingly unlikely
for models in which more representations that are added.

The above restrictions are enough to give a predictive scheme that
covers a large class of models. The origin of the magnitude of $M_I$
or the quantum number assignments of the extra matter are dependent on
the precise model of the string theory and so we do not consider these
points in detail here. Possibilities for the generation of this scale
include string-type non-renormalisable operators
\cite{Dinesetal,originmassnonren} or operators generated by some
hidden sector dynamics \cite{GUST}.

We define the number of vector (3,1) representations to be $a$,
the number of vector (1,2) representations to be $b$,
and the number of vector (3,2) representations to be $c$,
where the vector representation corresponding to $(3,1)$ is
($(\bar{3},1)\oplus (3,1)$) and so on for the other representations.
We assume that each vector representation has an explicit mass $M_I$
so that the effect of the extra vector representations is felt in the
renormalisation scale region between $M_I$ and $M_X$. Below
$M_I$, the extra matter is integrated out of the effective field
theory which then becomes
the MSSM\@. The beta functions of the gauge couplings are defined by
\begin{equation}
16 \pi^2 \frac{\partial g^2_i}{\partial t} = - b_i g_i^4,
\label{RGB}
\end{equation}
where $t=\ln \mu_0^2 / \mu^2$, $g_i$ is the $i^{th}$ gauge coupling and
$\mu_o, \mu$ are the initial (high) and final (low)
$\overline{MS}$ renormalisation scales respectively.
The beta functions of the effective theory between $M_X$ and $M_I$ are
\begin{equation}
b_i = \left( \left[ 11 + G_{t} 
\right] / k_1, \left[ 1 + b +3c \right], \left[ -3 + a + 2c \right]
\right)
\label{Bfns}
\end{equation}
where $k_1$ is the string normalisation of the U(1)$_Y$ gauge coupling
defined by
\begin{equation}
\alpha_1^{string} = k_1 \alpha_Y^{SM}, \label{stringnorm}
\end{equation}
$\alpha_Y^{SM}$ being the
hypercharge gauge coupling in the standard model normalisation. 
A particular example $k_1=5/3$ yields the standard GUT normalisation
\begin{equation}
\alpha_1^{GUT} = (5/3) \alpha_Y^{SM}. \label{GUTnorm}
\end{equation}
We have also used 
$G_t \equiv \Sigma_i (Y_i/2)^2$ where $i$  runs over all of the
superfields
additional to the MSSM with Standard Model hypercharges $Y_i/2$. 

In order to make the calculation as general as possible, model
dependent factors such as string threshold corrections are not
included. Given
this substantial approximation, it is sufficient to use first
order perturbation theory, a degenerate SUSY spectrum with mass
$M_{SUSY}$ and the step function approximation for mass thresholds in
the renormalisation group (RG) equations. Using these approximations we
obtain
\begin{eqnarray}
{\alpha_1(M_X)}^{-1} &=& \frac{5}{3k_1} \left( {\alpha_1(M_Z)}^{-1} 
+ \frac{53}{30 \pi} \ln M_Z +\frac{17}{60\pi} \ln m_t  + \frac{5}{4\pi}
\ln M_{SUSY} + \right. \nonumber \\&&\left. \frac{3}{10\pi}G_t \ln M_I-
\frac{3}{10\pi}(11+G_t) \ln M_X 
\right) \label{alpha1}\\
{\alpha_2(M_X)}^{-1} &=&  {\alpha_2(M_Z)}^{-1} 
- \frac{25}{12 \pi} \ln M_Z + \frac{1}{2\pi} \ln m_t + \frac{25}{12\pi}
\ln M_{SUSY} + \nonumber \\&&\frac{1}{2\pi} p \ln M_I- \frac{1}{2\pi}(1+p)
\ln M_X \label{alpha2}\\ 
{\alpha_3(M_X)}^{-1} &=&  {\alpha_3(M_Z)}^{-1} 
- \frac{23
}{6 \pi} \ln M_Z + \frac{1}{3\pi} \ln m_t + \frac{2}{\pi}
\ln M_{SUSY} + \nonumber \\
&&\frac{1}{2\pi}(a+2c)\ln M_I+\frac{1}{2\pi} (3 - a-2c)\ln M_X
\label{alpha3} 
\end{eqnarray}
where the $\alpha_i(M_X)\equiv \alpha_i^{string}(M_X)$ 
are all in the string normalisation and
${\alpha_1(M_Z)}^{-1}\equiv {\alpha_1^{GUT}(M_Z)}^{-1} = 58.89$ 
is in the GUT normalisation. We have
defined the positive integer $p=b+3c$, which counts the number of
additional SU(2)$_L$ doublets. Note that all of the mass scales
referred to in this paper are $\overline{MS}$ running masses, except
pole masses which are denoted with a superscript $phys$. For example,
to first order, the top quark pole mass is related to its running mass
by 
\begin{equation}
m_t^{phys} =
m_t \left[ 1 + 4 \alpha_s \frac{m_t}{3 \pi} \right].
\end{equation}

The normalisation $k_1$ is very model dependent, the most general
constraint upon it is that it must be rational and greater than or
equal to
one \cite{ibanez,nonst}. We
therefore regard it as a
free parameter and so have one less prediction, that of
$\sin^2\theta_W$,
compared to SUSY GUTs. Eq.\ref{stringscale} partly compensates for this by a
prediction of the string scale $M_X$ in terms of any of the gauge couplings.
Once $M_X$ is determined, the left hand sides of
Eqs.\ref{alpha2},\ref{alpha3} may be equated to
yield a prediction for the intermediate scale
\begin{eqnarray}
\ln\frac{M_I}{M_X} &=& \frac{1}{n} \left[4\ln M_X - \left(2 \pi (
{\alpha_2(M_Z)}^{-1} - {\alpha_3(M_Z)}^{-1}) + \frac{7}{2} \ln M_Z +
\right. \right. \nonumber \\&& \left. \left.
\frac{1}{3} \ln m_t + \frac{1}{6} \ln M_{SUSY} \right) \right],
\label{pred1}
\end{eqnarray}
where the integer $n$ is defined by $n\equiv b+c-a$. Eq.\ref{pred1}
allows us to get a 
bound on $n$ by applying the constraint $M_I <  M_X$. Using the input
parameters ${\alpha_2(M_Z)}^{-1} = 29.75$,
$\alpha_3(M_Z)=0.112-0.122$, $m_t^{phys}=152-196$ GeV\footnote{The top
quark mass
measurement by the CDF collaboration~\cite{CDF}.}, $M_X = 3-5 \times
10^{17}$ GeV and $M_{SUSY}=200-1000$ GeV, we obtain that the quantity
within the square brackets on the right hand side of Eq.\ref{pred1} is
always positive and hence $n<0$, or
\begin{equation}
a>b+c. \label{nbound} \end{equation}
All of the possible MSSM+X models satisfying Eq.\ref{nbound} with up
to 5 vector representations added are displayed in
Table~\ref{tab:models}. It is upon these simple examples that we shall
focus our attention.
\begin{table}
\begin{center}
\begin{tabular}{|c|cccccc|} \hline
Model&$N$&$a$&$b$ & $c$ & $n$ & $p$ \\ \hline
A & 1 & 1 & 0 & 0 & -1 & 0\\
B & 2 & 2 & 0 & 0 & -2 & 0\\
C & 3 & 2 & 1 & 0 & -1 & 1\\
D & 3 & 2 & 0 & 1 & -1 & 3\\
E & 3 & 3 & 0 & 0 & -3 & 0\\
F & 4 & 3 & 1 & 0 & -2 & 1\\
G & 4 & 3 & 0 & 1 & -2 & 3\\
H & 4 & 4 & 0 & 0 & -4 & 0\\
I & 5 & 3 & 1 & 1 & -1 & 4\\
J & 5 & 3 & 2 & 0 & -1 & 2\\
K & 5 & 3 & 0 & 2 & -1 & 6\\
L & 5 & 4 & 1 & 0 & -3 & 1\\
M & 5 & 4 & 0 & 1 & -3 & 3\\
N & 5 & 5 & 0 & 0 & -5 & 0\\ \hline
X & 25 & 14 & 10 & 1 & -3 & 13\\ \hline
\end{tabular}
\end{center}
\caption{MSSM+X models with $N\leq5$ additional 
chiral superfields in vector
representations of the MSSM that satisfy Eq.\protect\ref{nbound}. The
7 columns detail the names and content of the models,
where $a,b,c$ are the number of chiral scalar fields
in the vector rep.s $(3,1),(1,2),(3,2)$, respectively,
and $n=b+c-a$, $p=b+3c$.
The final row details a special
model containing $N=25$ additional superfields, 
which is introduced for the purposes of the discussion
in section 4.} 
\label{tab:models}
\end{table}

For the case of many extra gauge representations to the MSSM with mass
$\sim M_I$, another bound may be placed upon $a,b,c$ as follows.
Eq.s~\ref{alpha2},\ref{pred1},\ref{stringscale} may be rearranged 
such that the ratio $p/n$ may be expressed in
terms of ${\alpha_2(M_X)}^{-1}$:
\begin{equation}
\frac{p}{n} = \frac{2 \pi {\alpha_2(M_X)}^{-1} - \frac{1}{2} \ln
\left( \frac{{\alpha_2(M_X)}^{-1} }{4 \pi} \right) + \ln Z - 2 \pi X}
{4 \ln Z  - 2 \ln \left( \frac{{\alpha_2(M_X)}^{-1} }{4 \pi} \right)
- Y}, \label{pon}
\end{equation}
where we have defined
\begin{eqnarray}
X &\equiv & {\alpha_2(M_Z)}^{-1} - \frac{25}{12 \pi} \ln M_Z +
\frac{25}{12 \pi} \ln M_{SUSY} + \frac{1}{2 \pi} \ln m_t \nonumber \\
Y &\equiv& 2 \pi \left( {\alpha_2(M_Z)}^{-1} - {\alpha_3(M_Z)}^{-1}
\right) + \frac{7}{2} \ln M_Z + \frac{1}{3} \ln m_t + \frac{1}{6} \ln
M_{SUSY} \nonumber \\
Z &=& 5.3 \times 10^{17} \mbox{~GeV}. \nonumber 
\end{eqnarray}
The right hand side of Eq.\ref{pon} is rather complicated
but can be investigated numerically as a function
of ${\alpha_2(M_X)}^{-1}$. We find that $p/n$ has a minimum as a function of 
${\alpha_2(M_X)}^{-1}$, albeit with a 
large uncertainty from $\alpha_3 (M_Z)$. When the minimum of Eq.\ref{pon} is
determined numerically, we obtain the bound
\begin{equation} p/n > K,
\end{equation} where $K =
-11.0,-9.2,-8.0$ for $\alpha_S(M_Z)=0.112,0.117,0.122$ respectively.
Since $n$ must be negative, and $p$ is positive,
the bound may be written as 
\begin{equation} p<|n||K|,
\end{equation}
from which we see that the number of doublets $p$ is bounded from above.
This bound is not approached for the models in Table 1, but will
be relevant when we come to consider the origin of the Yukawa matrices
in section 4.

\begin{figure}
\begin{center}
\leavevmode
\hbox{%
\epsfxsize=4.5in
\epsfysize=3in
\epsffile{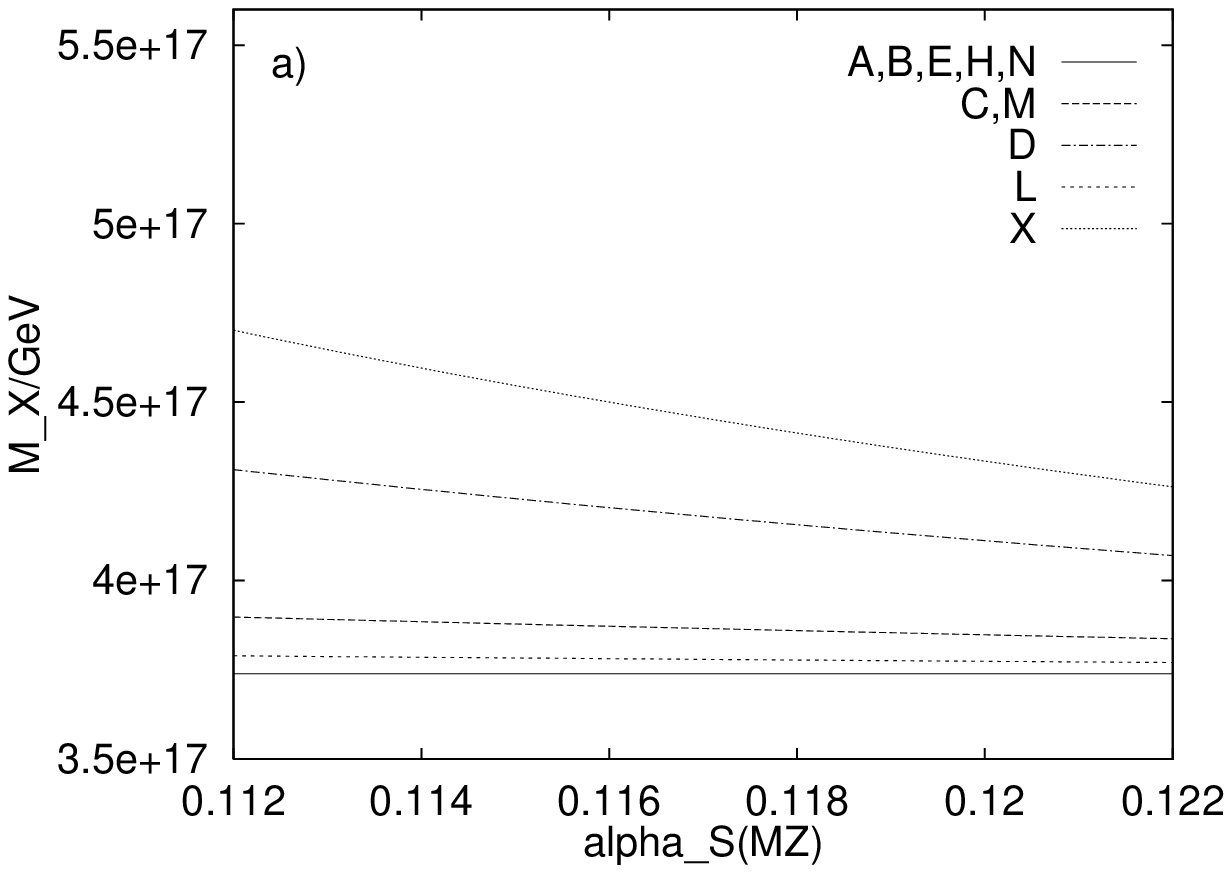}}
\vspace{\baselineskip}

\leavevmode
\hbox{%
\epsfxsize=4.5in
\epsfysize=3in
\epsffile{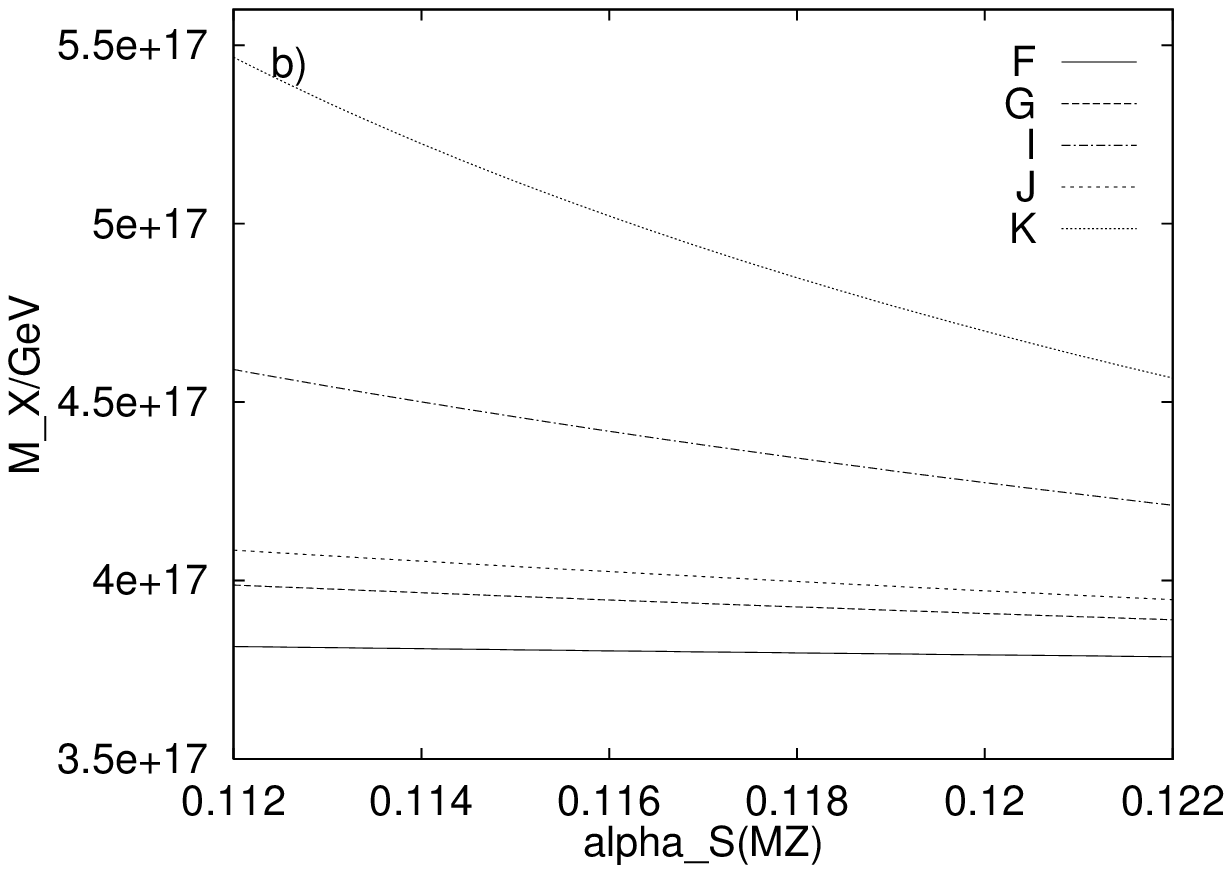}}
\end{center}
\caption{Prediction of the string scale $M_X$ in the MSSM+X models for
$M_{SUSY}=500$ GeV, $m_t^{phys}=174$ GeV and various $\alpha_S(M_Z)$. 
The key identifies the model by reference to
Table~\protect\ref{tab:models}.}
\label{fig:MX}
\end{figure}
To extract $M_X$ in practice requires an iterative numerical
procedure. First a scale $\Lambda \sim 5.10^{17}$ GeV is substituted
for $M_X$ in Eq.\ref{pred1} to give an $M_I$ consistent with gauge
unification at a scale $\Lambda$. This value of $M_I$ is then
substituted into Eq.\ref{alpha2} to yield ${\alpha(\Lambda)}^{-1}$.
Substituting ${\alpha(\Lambda)}$ into Eq.\ref{stringscale} yields what
string breaking scale $M_X$ would correspond to this gauge coupling.
This $M_X$ is substituted for $\Lambda$ and the whole process is
repeated until $\Lambda=M_X$ is converged upon.
If this procedure converges, we are left with numerical values of
$M_I, M_X, \alpha_i(M_X)$ that are consistent with gauge unification
in the model under study.

\begin{figure}
\begin{center}
\leavevmode
\hbox{%
\epsfxsize=4.5in
\epsfysize=3in
\epsffile{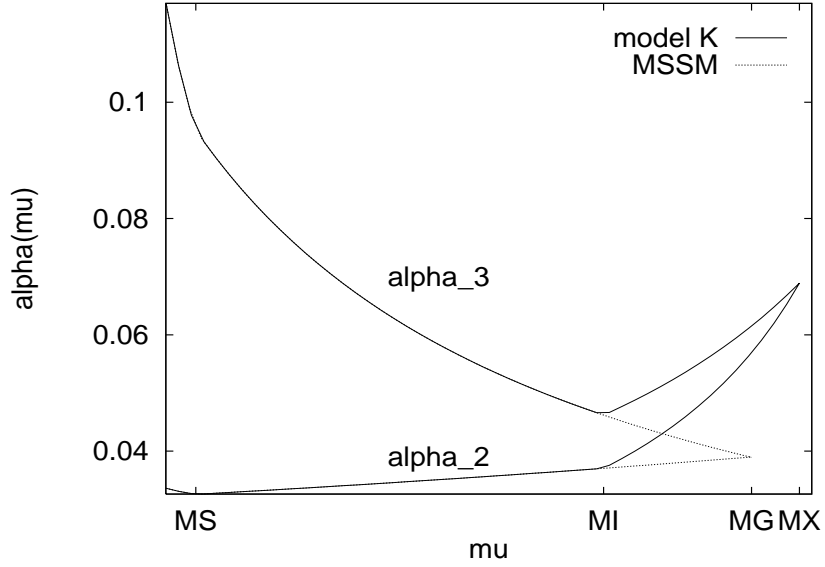}}
\end{center}
\caption{Comparison between model K and the MSSM of the running of the
gauge coupling constants $\alpha_2,\alpha_3$. 
The running is between $\mu=M_Z$ and $\mu=M_X, M_G=10^{16}$ GeV for
$m_t^{phys}=174$ GeV, 
$M_{SUSY}=500$
GeV and $\alpha_S(M_Z)=0.117$.}
\label{fig:gauge1}
\end{figure}

\begin{figure}
\begin{center}
\leavevmode
\hbox{%
\epsfxsize=4.5in
\epsfysize=3in
\epsffile{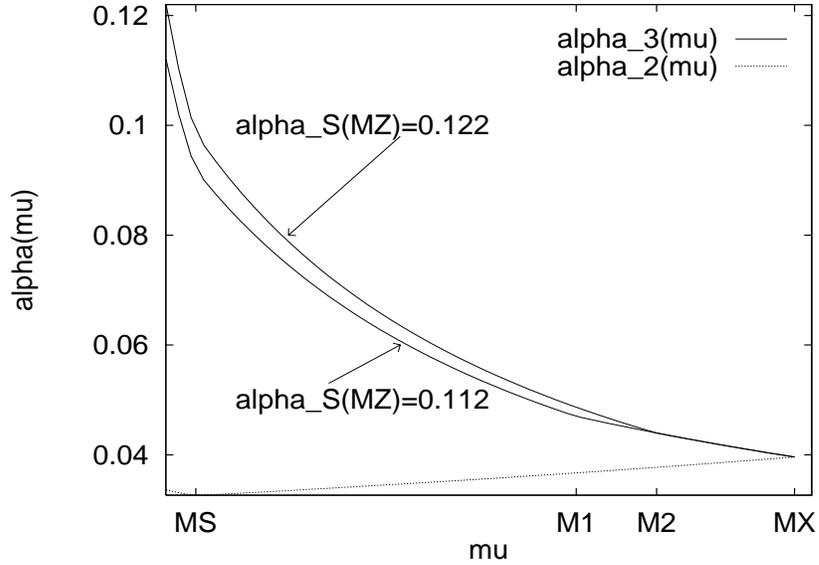}}
\end{center}
\caption{Running of the SU(3) and SU(2)$_L$ gauge coupling constants
in model A
between $M_Z$ and $M_X$ for $m_t^{phys}=174$ GeV, $M_{SUSY}=500$ GeV.
$M_I
\equiv M1,M2$
correspond to $\alpha_S(M_Z)=0.112,0.122$ respectively.}
\label{fig:gauge2}
\end{figure}

Fig.\ref{fig:MX} displays the values of $M_X$ given by this procedure
for the models outlined in Table~\ref{tab:models}. Central
values of $m_t, M_{SUSY}$ were picked and the results have negligible
sensitivity upon $m_t$. Varying $M_{SUSY}$ between 200 and 1000 GeV changes
the $M_X$ prediction by $\sim 0.1 \times 10^{17}$ GeV. As shown, the
results are in general quite dependent upon the strong coupling
constant $\alpha_S (M_Z)$ and so we have used this as the independent
parameter in the plots. In Fig.\ref{fig:gauge1} the
running of the gauge couplings in model K is compared to the running
purely within the MSSM\@. At $M_I$, the effects of the extra
representations are felt and $\alpha_{2,3}$ rise steeply with $\mu$.
The
general tendency shown by Fig.\ref{fig:MX} is that $M_X$ is higher for
models which possess the most SU(2)$_L$ doublets (high $p$), and
lower for models
in which the number of SU(3) triplets minus the number of SU(2)$_L$
doublets is great (more negative $n$). The class $p=0$ when there are no
added doublets for models A,B,E,H,N is a special class of cases in
which there
is no $\alpha_S(M_Z)$ dependence to 1 loop. 
Because there are no more SU(2) representations
than the MSSM, the running of $\alpha_2$ is identical to the MSSM
until the GUT scale $M_{G}$. This alone fixes $M_X, \alpha_i(M_X)$ in
these cases, since for
$p=0$, Eq.\ref{stringscale} and Eq.\ref{alpha2} could be combined to
give an equation with only one output: $M_X$ for example. 
An example of this case is
model A ($n=-1,p=0$), which is examined in more detail in
Fig.\ref{fig:gauge2}. It is shown in the figure that when different
initial values of $\alpha_S(M_Z)$ are taken, $M_I$
conspires to give the same value of $M_X$ (and therefore
$\alpha_i(M_X)$). 

\begin{figure}
\begin{center}
\leavevmode
\hbox{%
\epsfxsize=4.5in
\epsfysize=3in
\epsffile{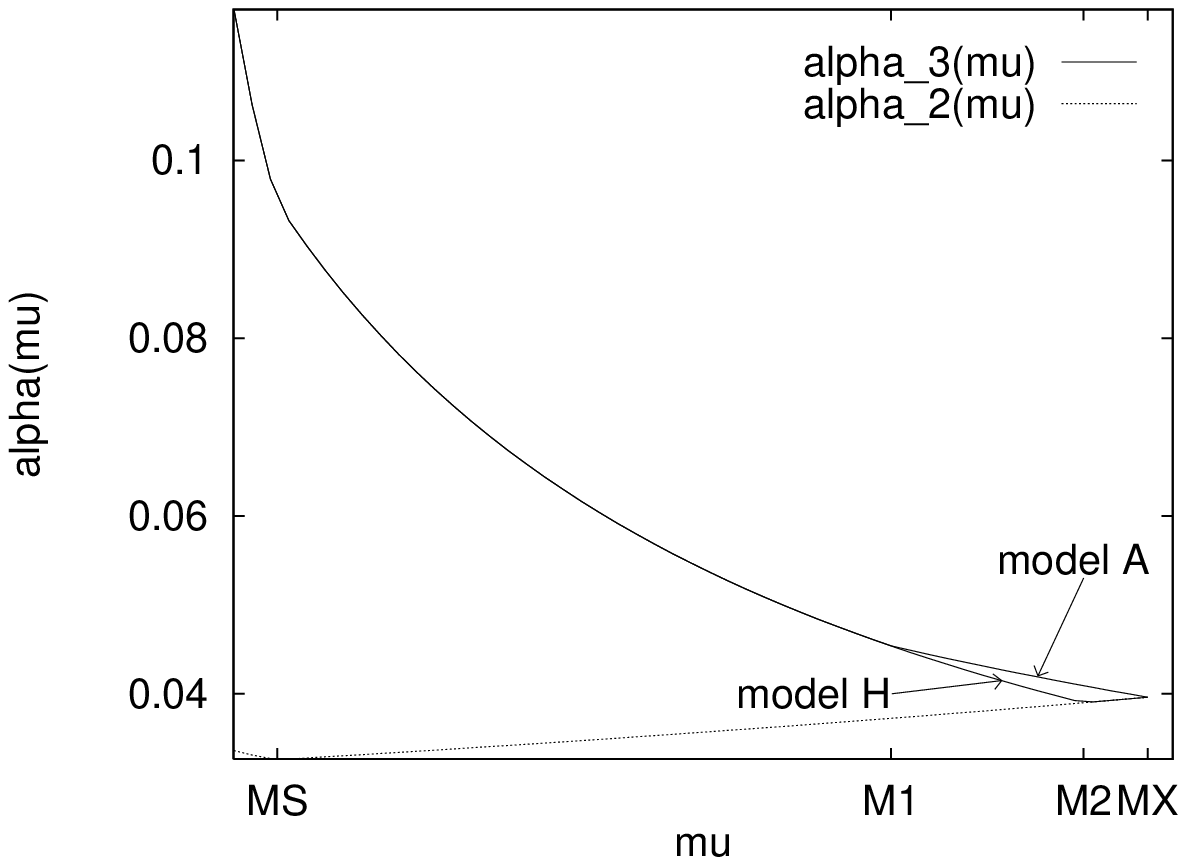}}
\end{center}
\caption{Comparison of the running of the gauge coupling constants in
model A and model H.
The running is between $\mu=M_Z$ and $\mu=M_X$ for $m_t^{phys}=174$,
GeV
$M_{SUSY}=500$ GeV and $\alpha_S(M_Z)=0.117$. $M_I\equiv M1,M2$
correspond to
models A,H respectively.}
\label{fig:gauge3}
\end{figure}
\begin{figure}
\begin{center}
\leavevmode
\hbox{%
\epsfxsize=4.5in
\epsfysize=3in
\epsffile{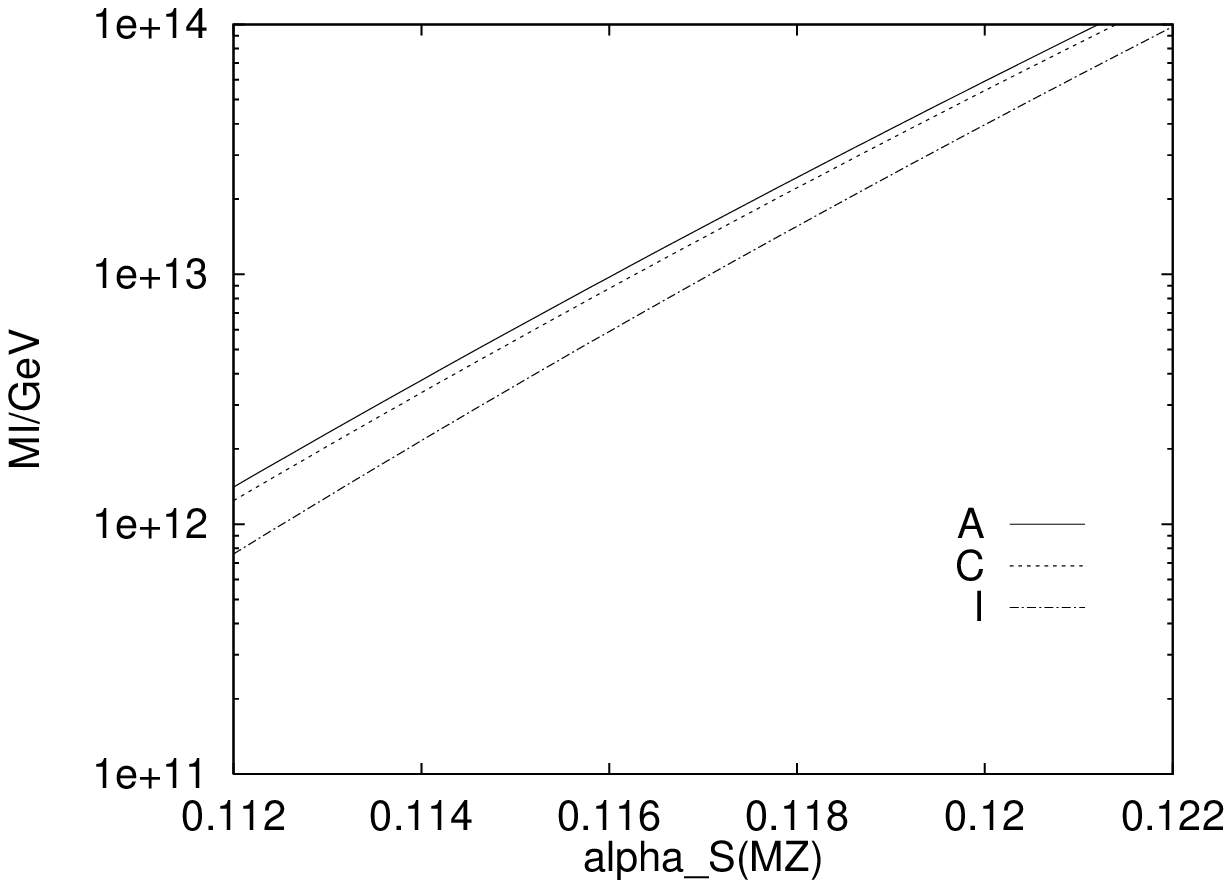}}\end{center}
\caption{Prediction of the intermediate scale $M_I$ in the MSSM+X
models indicated in the key for 
$M_{SUSY}=500$ GeV, $m_t^{phys}=174$ GeV and various $\alpha_S(M_Z)$.}
\label{fig:MI1}
\end{figure}
\begin{figure}
\begin{center}
\leavevmode
\hbox{%
\epsfxsize=4.5in
\epsfysize=3in
\epsffile{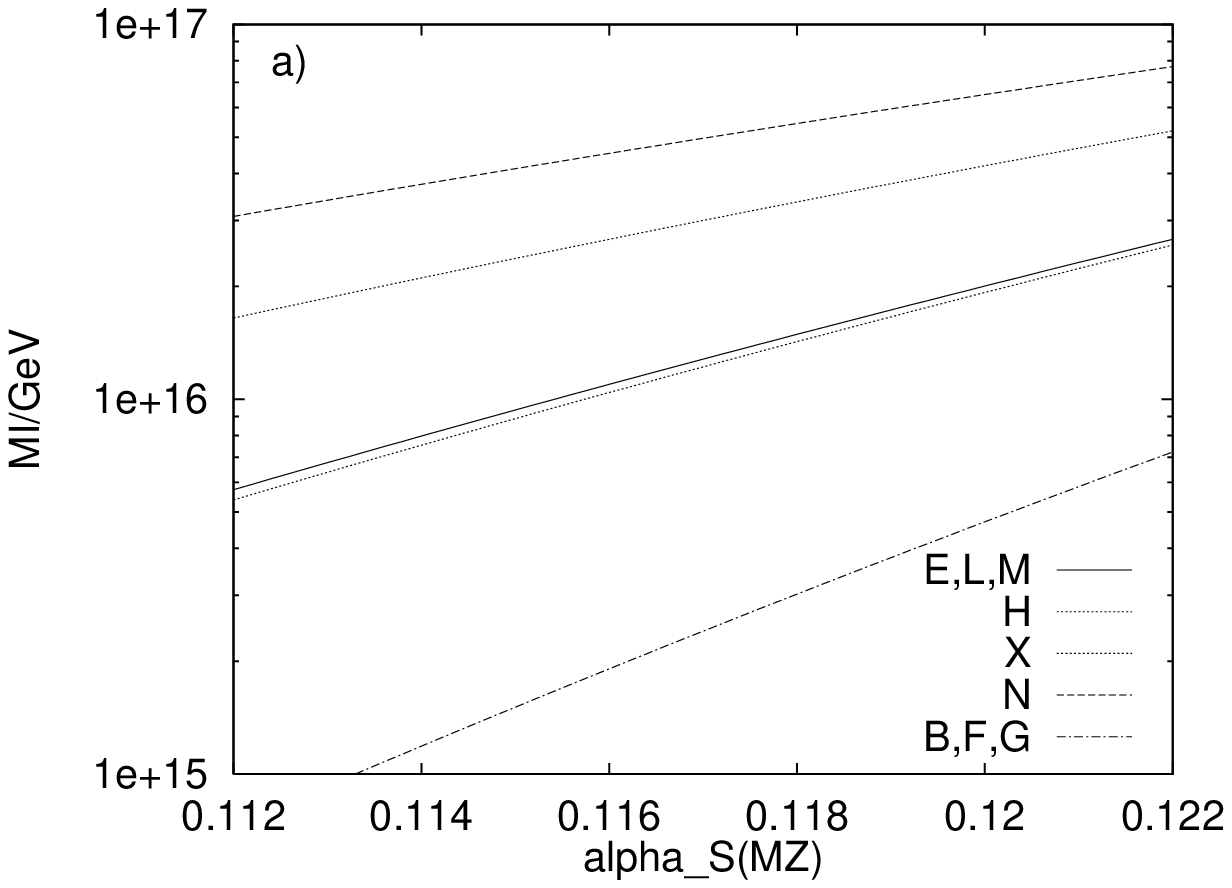}}
\vspace{\baselineskip}

\leavevmode
\hbox{%
\epsfxsize=4.5in
\epsfysize=3in
\epsffile{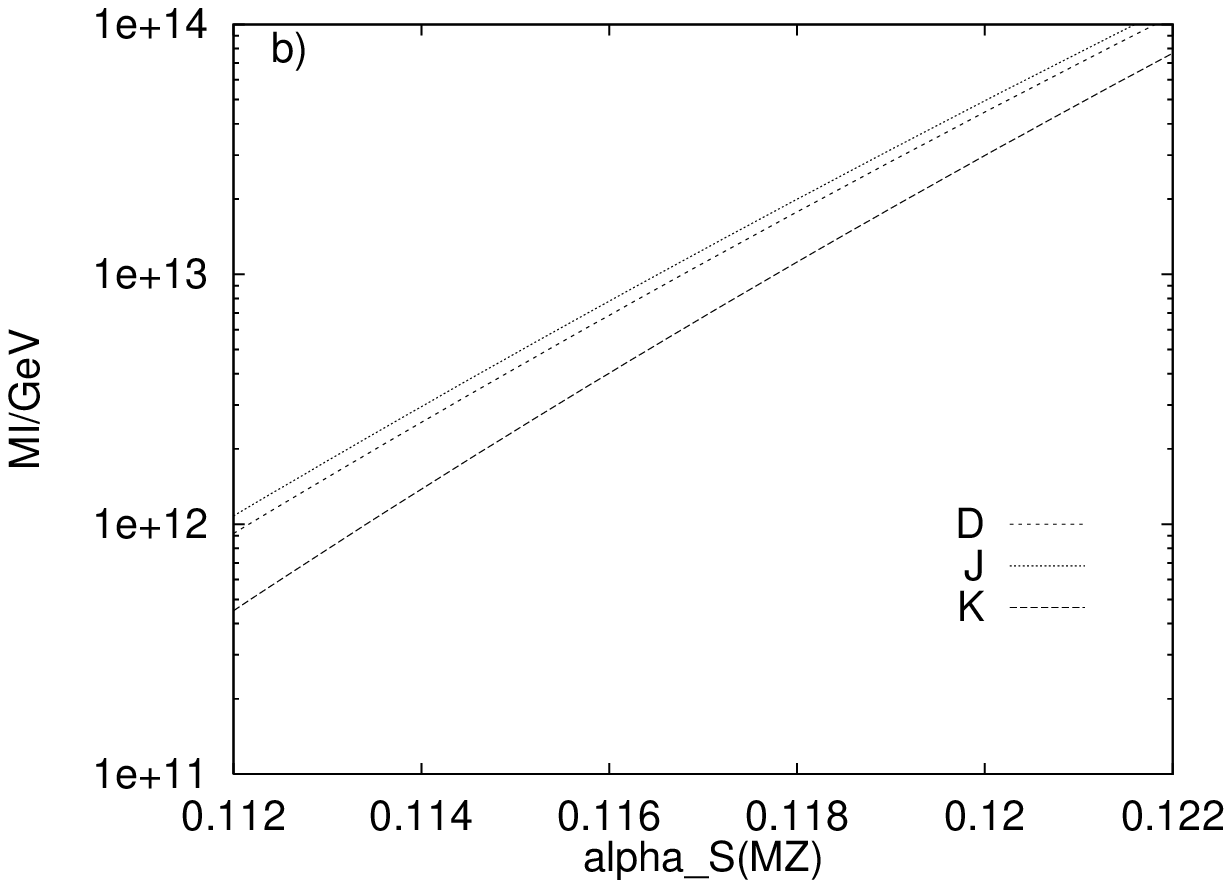}}
\end{center}
\caption{Prediction of the intermediate scale $M_I$ in the MSSM+X
models indicated in the key for
$M_{SUSY}=500$ GeV, $m_t^{phys}=174$ GeV and various $\alpha_S(M_Z)$.}
\label{fig:MI2}
\end{figure}

Figs.\ref{fig:MI1},\ref{fig:MI2} show the predictions for $M_I$ for
each of the
models $A,B,\ldots,N$. Varying $M_{SUSY}$ between 200 and 1000 GeV makes
a maximum difference to $\ln M_I$ of $\sim 2\%$ and the results (like
all of the
gauge unification predictions) are not very
dependent on $m_t^{phys}=152-196$ GeV. The results are dependent upon the
value of $n$ that is relevant for the model in question. This is
because $n$ counts the number of extra SU(2)$_L$ doublets minus the number
of extra SU(3) triplets in the model. This point is illustrated in
Fig.\ref{fig:gauge3}, where model A ($n=-1, p=0$) is compared with
model H ($n=-4,p=0$). Models with $p=0$ have $M_X, \alpha_i(M_X)$ fixed
independent of $n$ as stated previously and models with higher $-n$ have
more positive slopes in the region $M_I < \mu < M_X$. Thus, to hit the
same endpoint of $\mu=M_X,\alpha_i(\mu)$, the lower $-n$ models must
have lower $M_I$ in order to agree with the low energy gauge
couplings.

\begin{figure}
\begin{center}
\leavevmode
\hbox{%
\epsfxsize=4.5in
\epsfysize=3in
\epsffile{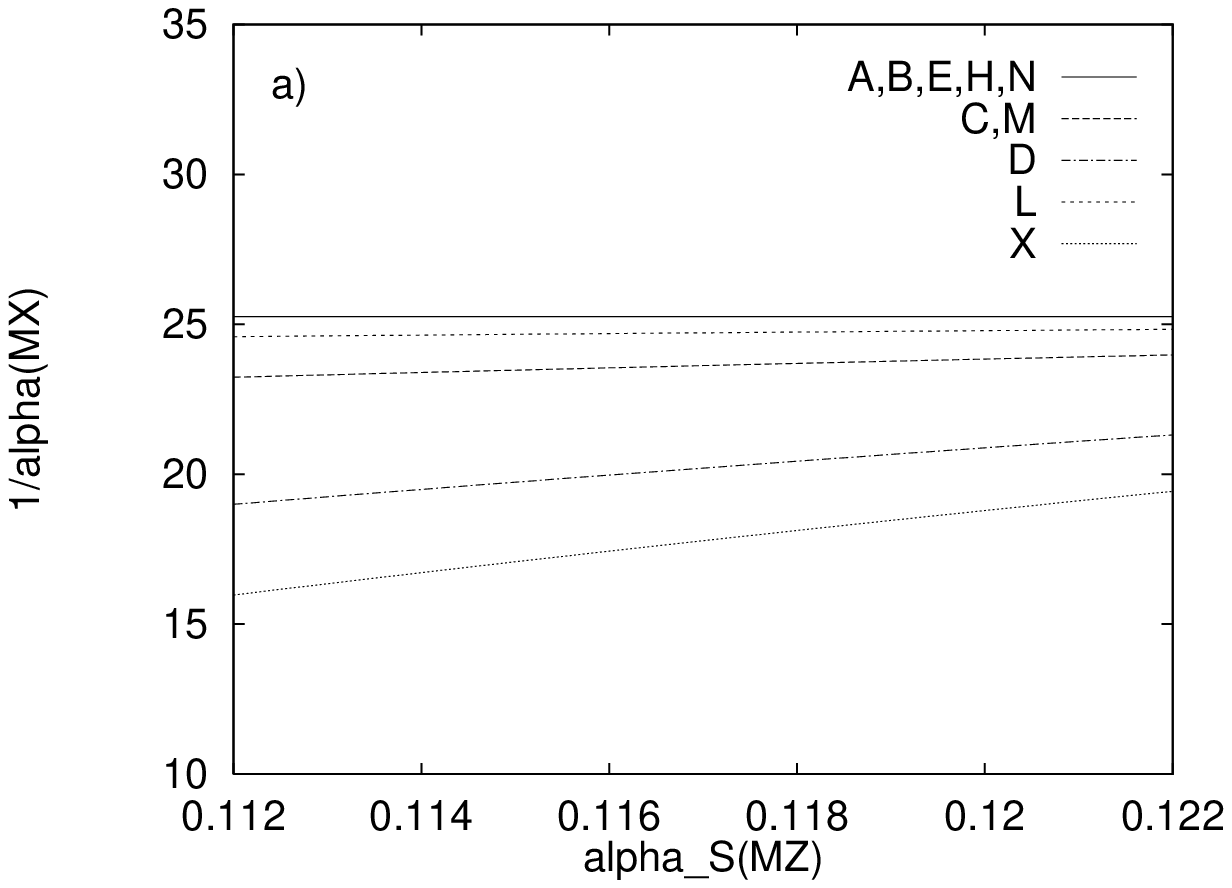}}
\vspace{\baselineskip}

\leavevmode
\hbox{%
\epsfxsize=4.5in
\epsfysize=3in
\epsffile{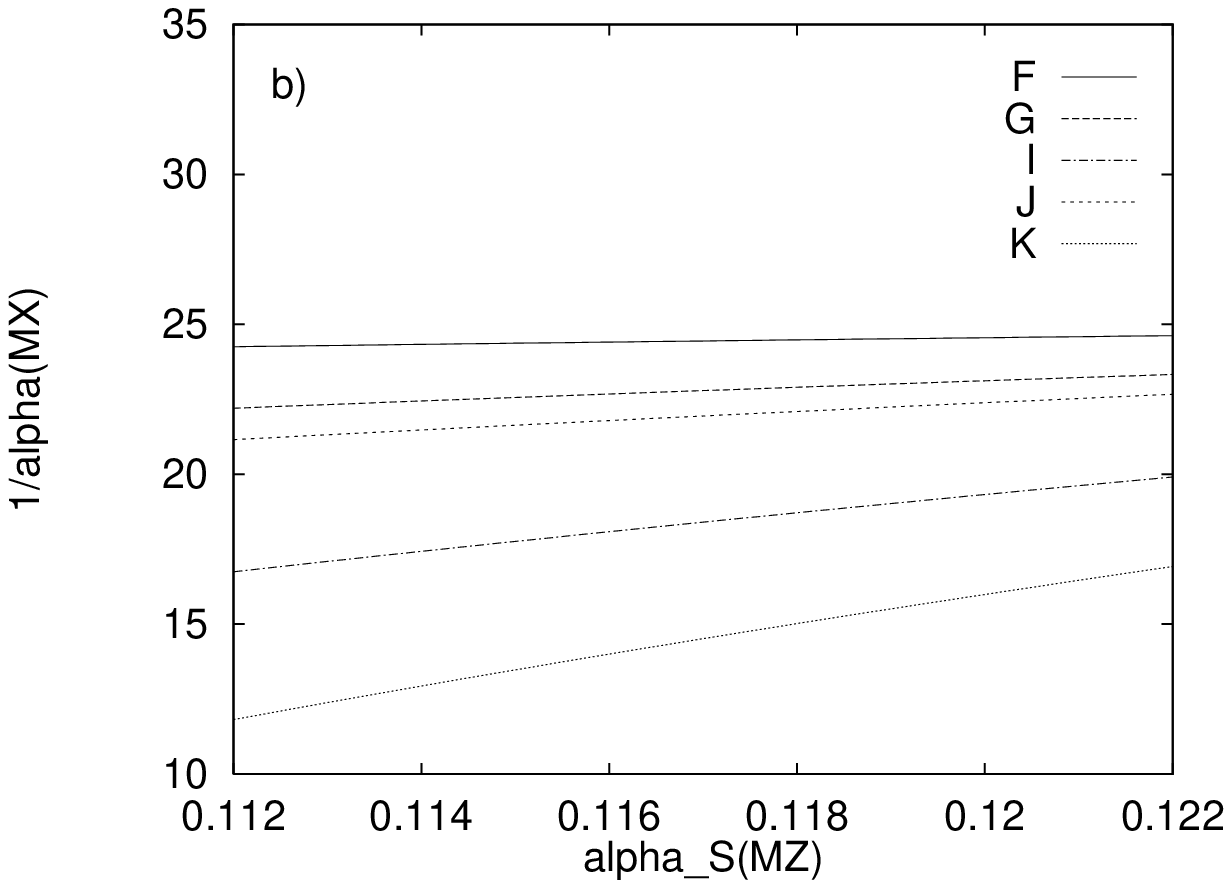}}
\end{center}
\caption{Prediction of the string gauge coupling ${\alpha(M_X)}^{-1}$
in the MSSM+X models for
$M_{SUSY}=500$ GeV, $m_t^{phys}=174$ GeV and various $\alpha_S(M_Z)$. 
The key identifies the model by reference to
Table~\protect\ref{tab:models}.}
\label{fig:aGUT}
\end{figure}

The predictions of $\alpha_i(M_X)$ vary a lot depending upon how many
SU(2)$_L$ doublets are present in the intermediate region $M_I < \mu <
M_X$, as shown by Fig.\ref{fig:aGUT}. 
Models with high values of $p$ tend to have high $\alpha_i(M_X)$
because $b_{2,3}$ (and therefore the rates of change of the gauge
couplings with respect to $\mu$) are more positive, as Fig.\ref{fig:gauge1}
shows. $\alpha_i(M_X)$ is approximately not dependent upon $m_t$ and
$M_{SUSY} = 200-1000$ GeV corresponds to a change in $\alpha_i(M_X)$ of
$\sim 2\%$.

\begin{figure}
\begin{center}
\leavevmode
\hbox{%
\epsfxsize=4.5in
\epsfysize=3in
\epsffile{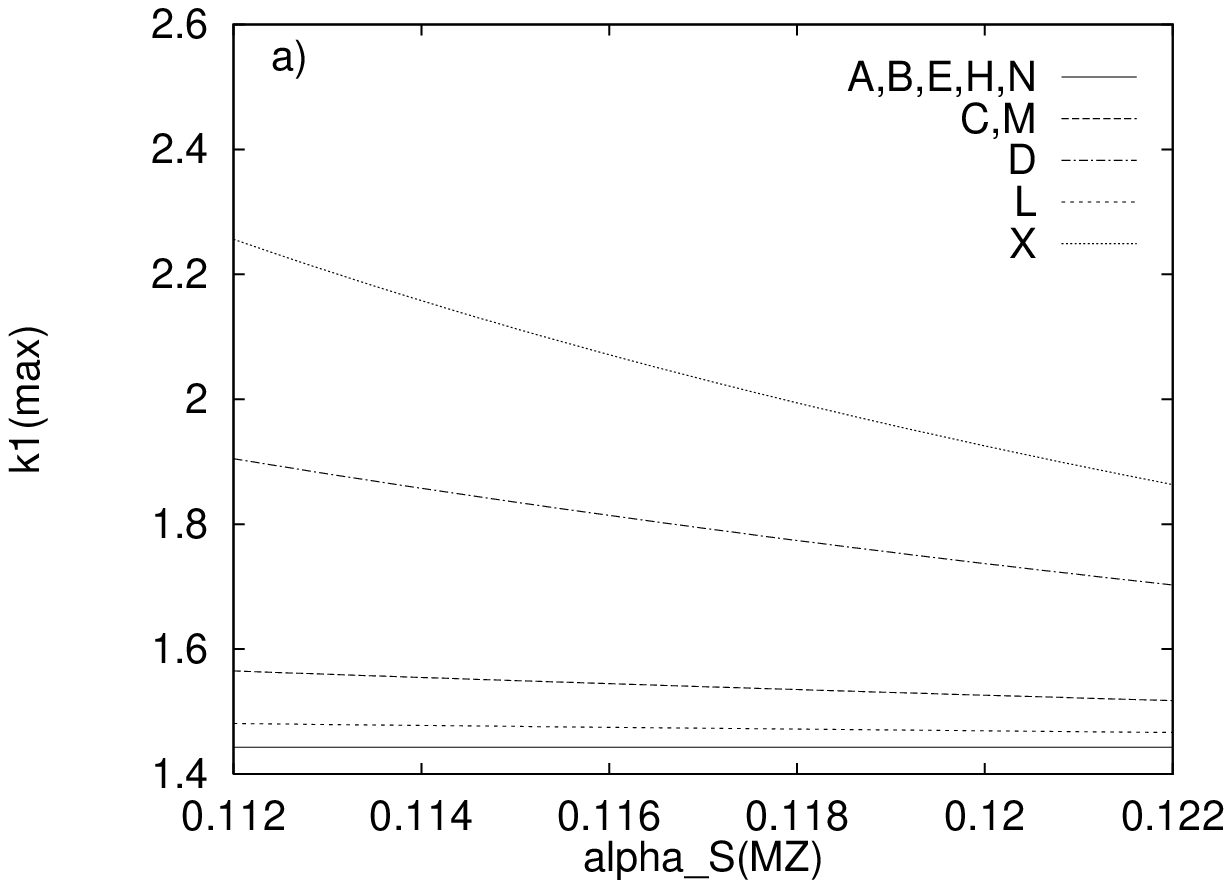}}
\vspace{\baselineskip}

\leavevmode
\hbox{%
\epsfxsize=4.5in
\epsfysize=3in
\epsffile{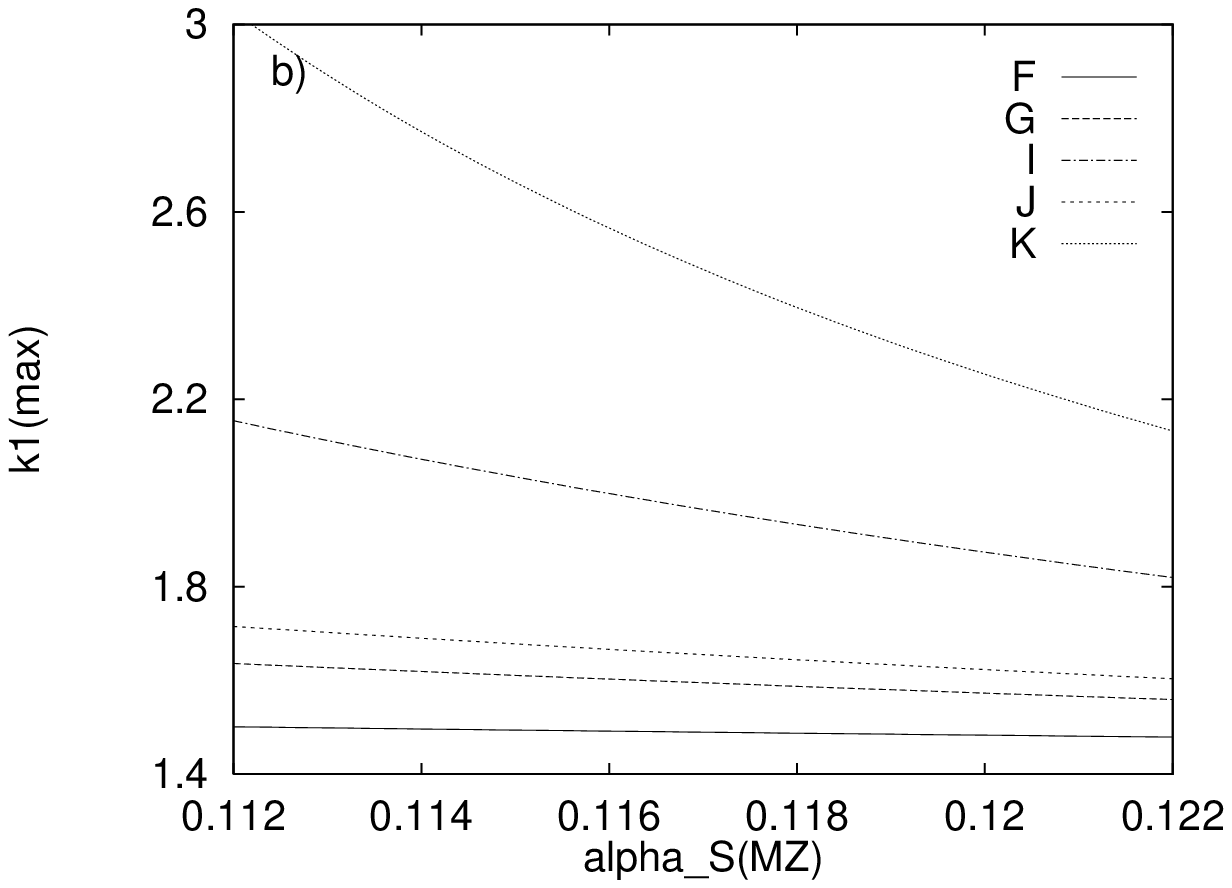}}
\end{center}
\caption{The maximum value of $k_1$
in the MSSM+X models for
$M_{SUSY}=500$ GeV, $m_t^{phys}=174$ GeV and various $\alpha_S(M_Z)$. 
The key identifies the model by reference to
Table~\protect\ref{tab:models}.}
\label{fig:k1max}
\end{figure}

We may now extract some more information about the string theory from
which the MSSM+X models must derive by examining the unification of
the hypercharge gauge coupling. Setting the right hand sides of
Eqs.\ref{alpha1},\ref{alpha2} equal yields an equation
\begin{equation}
k_1 = \frac{5}{3} \frac{2 \pi {\alpha_1(M_Z)}^{-1} + \frac{53}{15}
\ln M_Z + \frac{17}{30} \ln m_t + \frac{5}{2} \ln M_{SUSY} - \frac{3
G_t}{5} \ln \left(\frac{M_X}{M_I}\right) - \frac{33}{5} \ln M_X}
{2 \pi {\alpha_2(M_Z)}^{-1} - \frac{25}{6}
\ln M_Z + \ln m_t + \frac{25}{6} \ln M_{SUSY} + p\ln M_I - (1+p) \ln M_X}.
\label{k1pred}
\end{equation}
Eq.\ref{k1pred} cannot be used to predict the string normalisation
$k_1$ since $G_t$ is arbitrary and unknown. However, an upper bound
may be placed upon $k_1$ by noting that $G_t$ is positive
semi-definite. Setting $G_t=0$ in Eq.\ref{k1pred} therefore gives the
maximum string
normalisation upon the hypercharge assignments consistent with gauge
unification at the string scale, and therefore placing that constraint
upon the string theory that is supposed to reduce to the MSSM+X model
as the low energy
effective field theory limit.
Fig.\ref{fig:k1max} displays the upper bounds upon $k_1$ for the
MSSM+X models A,B,\ldots,N. Higher $p$ and lower $-n$ corresponds to a
higher upper
bound, mostly because in these cases $\alpha_i(M_X)$ is large, as
explained earlier.
Again, the results are roughly independent of $m_t$ and only depend on
$M_{SUSY}$ at the $\sim 2\%$ or less level. As an example, the only MSSM+X
models studied here that are consistent with the GUT normalisation of
$k_1=5/3$ are D,I,J,K,X.
The bound $k_1 \geq 1$ \cite{ibanez} may be used in Eq.\ref{k1pred} to
place an
upper bound upon $G_t$. As Table~\ref{tab:Gtot} shows, the maximum
hypercharge assignments for the extra matter are
large compared with typical hypercharges in the Standard
Model. The numbers are so large that they are unlikely to be a strong
constraint on a given string model (for the whole of the MSSM, $\sum_i
(Y_i/2)^2 = 11$). Once $k_1$ is picked in the context of a particular
string model, $G_t$ is then fixed.
\begin{table}
\begin{center}
\begin{tabular}{|c|cccccccccccccc|} \hline
$\alpha_S(M_Z)$&A&B&C&D&E&F&G&H&I&J&K&L&M&N \\ \hline
0.112&5.7&11&16.5&8&17&12&14&22&9&
7&11&18&20&28\\
0.122&9.1&18&9.8&11&27&19&20&36&12&11&14&28&29&45\\ \hline
\end{tabular}
\end{center}
\caption{Upper bounds on $G_t$ for $M_{SUSY}=200-1000$ GeV and
$m_t^{phys}=152-196$ GeV for the MSSM+X models in
Table~\protect\ref{tab:models}.}
\label{tab:Gtot}
\end{table}
As an example of what possible hypercharge normalisations may result,
we focus on the particular example of model D, which is
equivalent to the MSSM plus 2 right handed quark representations and
one left handed quark representation at the scale $M_I\sim10^{12-14}$
GeV. Assuming these superfields have the same Standard Model
hypercharge assignments as $Q_L, u_R, d_R$ respectively, we obtain
$k_1 \sim 5/3$.

\section{Infra-red Fixed Points in MSSM+X Models \label{sec:FP}}

Lanzagorta and Ross \cite{RLfixed}
recently revisited the fixed
point \cite{PR,H}
in the RGEs of the top quark Yukawa coupling and QCD gauge
coupling in the framework of MSSM and SUSY GUT models.
In this section we shall extend their analysis to the
MSSM+X models
considered in the previous section.

The effective superpotential of the MSSM+X models is assumed to be
\begin{equation}
W = h_t Q H_2 u^c - \mu
H_1 H_2 + \ldots
\label{super}
\end{equation}
where $h_t$ is the top Yukawa coupling, $Q, u^c$ refer to the third
family left
handed quark and right handed quark superfields and $H_{1,2}$ are the
two Higgs doublet superfields. It has been assumed in Eq.\ref{super}
that the ratio of Higgs vacuum expectation values (VEVs)
$\tan \beta \equiv v_2/v_1$ is of
order one and all small Yukawa couplings have been dropped. The terms
due to the extra matter are assumed to be all of the forms $M_I (3,1).
(\bar{3},1)$, $M_I (1,2).(1,\bar{2})$, $M_I (3,2).(\bar{3},\bar{2})$
where
the group indices are traced over. Thus, these terms would give the
extra matter a mass $M_I$ but no extra parameters would enter the one
loop
top quark Yukawa 
coupling renormalisation group equation compared to the MSSM\@. Note
that this could be a
consequence of the extra matter having non-standard hypercharge
assignments\cite{nonst}, so that an additional superfield could not
couple to a
MSSM superfield with opposite SU(3)$\otimes$SU(2) quantum numbers
in a hypercharge invariant way.

The RGE for the case of only one large
Yukawa coupling is
\begin{equation}
\frac{\partial Y_t}{\partial t} = Y_t \left( \Sigma_i r_i
\tilde{\alpha}_i - s 
Y_t \right),
\label{YtRGE}
\end{equation}
where we have defined the parameters in the same notation as
Lanzagorta and Ross \cite{RLfixed} $\tilde{\alpha}_i \equiv g_i^2 / 16 \pi^2$,
$Y_t \equiv h_t^2 / 16 \pi^2$. 
Dropping the electroweak gauge couplings, Eq.\ref{YtRGE} can be written
as,
\begin{equation}
\frac{\partial R}{\partial t} = Y_t \left[(r_3+b_3) -sR\right]
\label{YtRGE2}
\end{equation}
where the ratio of Yukawa to gauge coupling has been written as

\begin{figure}
\begin{center}
\leavevmode
\hbox{%
\epsfxsize=3.15in
\epsfysize=1.82in
\epsffile{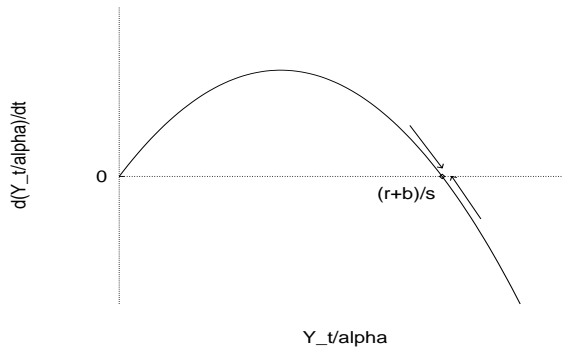}}
\end{center}
\caption{Infra red behaviour of $Y_t/\tilde{\alpha}$ in the case
when the top Yukawa coupling is dominant. The arrows indicate the
direction of flow for increasing $t$, i.e.\ in the direction of the
infra-red direction. The point
labeled $(r+b) / s$ is the fixed point.}
\label{fig:IRSFP}
\end{figure}

$R=\left(\frac{Y_t}{\tilde{\alpha}_3}\right)$.
For the MSSM, $r_3=16/3, b_3=-3, s=6$.
Eq.\ref{YtRGE2} has an infra-red stable fixed point 
given by 
\begin{equation}
R^*=\left( \frac{Y_t}{\tilde{\alpha}_3} \right)^*=(r_3+b_3)/s=7/18.
\end{equation}
as shown by Fig.\ref{fig:IRSFP}, where the asterisk denotes the fixed point.
The figure shows that $(Y_t/\tilde{\alpha})$ at an arbitrary scale
is attracted towards the fixed point as the energy scale is reduced.
The low energy value of $R=(Y_t/\tilde{\alpha})$ is given by
\begin{equation}
R(t)= \frac{R^*}
{1+ \Delta \left[ \frac{R^*}{R(0)} -1 \right]}
\label{FPclear}
\end{equation}
where we have defined
\begin{equation}
\Delta = \left( \frac{\alpha_3(t)}{\alpha_3 (0)} \right)^{B_3},
\label{Delta}
\end{equation}
$B_3 \equiv r_3/b_3 + 1$ and $t=\ln M_X^2/\mu^2$ 
where $\mu$ refers to the low energy
scale and $t=0$ corresponds to $\mu=M_{GUT}$.

There is no a priori reason why the high energy theory should
select the boundary condition on the top quark Yukawa coupling to be
at its fixed point.
However Eq.\ref{FPclear} shows that for arbitrary input value the
fixed point value is always reached in the limit $t \rightarrow
\infty$ since in this limit $\Delta \rightarrow 0$ (since
$B_3=-7/9$ and $\alpha_3$ is asymptotically free).
In practice $\Delta$ is finite ($\Delta \sim 0.5$ in the MSSM
from the running between GUT and weak scales)
and the low energy value of the top quark Yukawa coupling 
will be higher or lower than its fixed point value, depending on the
whether the high energy input value of the coupling is higher 
or lower than its fixed point value.
However it is well known that, for fixed $t$, there
is a maximum low energy value of top quark Yukawa coupling
corresponding to $Y_t(0)\rightarrow \infty$.
In fact this maximum value is achieved to very good accuracy by
finite but large input values which satisfy the condition
\begin{equation}
R(0) \gg R^*
\label{condition}
\end{equation}
which allows the simple approximate form of Eq.\ref{FPclear}
\begin{equation}
R(t)\approx \frac{R^*}{1 - \Delta}
\equiv R^{QFP}
\label{quasi}
\end{equation}
which is the well known quasi-fixed-point (QFP)\cite{H}.
It is worth emphasising that, given only that
the condition in Eq.\ref{condition} is met,
Eq.\ref{quasi} gives an (approximate) determination of the low energy
top quark Yukawa coupling which is quite insensitive to its 
high energy value. 
To be more specific, for any choice of input top Yukawa couplings
$R(0) > R^*$ the low energy values
of the top quark Yukawa couplings will be funneled into a
range between $R^*$ and $R^{QFP}$ where the
distance between the FP and QFP is controlled by the quantity $\Delta$
defined in Eq.\ref{Delta}.
The smaller the quantity $\Delta$, the closer will be the
QFP to the FP, and the more accurate will be the determination of the
low energy top Yukawa coupling.\footnote
{Parenthetically we note that the above analysis is not 
valid for the special case when $b_3=0$, so
that Eq.\ref{YtRGE} decouples from the running of the gauge coupling.
In this case, $Y^*=r\tilde{\alpha} / s$, with solution
$Y(t) = \frac{Y^*}{\frac{Y^*}{Y(0)} e^{-r \alpha t} - e^{- r \alpha
t} + 1}$
and the maximum distance from fixed point $\Delta = e^{- r \alpha t}$. 
Note that in a case where $b<0$ and $r<-b$, $Y_t \rightarrow 0$ at low
energy and has no infra-red fixed point since $\left(
\frac{\alpha_3(t)}{\alpha(0)} \right)^{B_3} \rightarrow \infty$ as $t
\rightarrow \infty$. This does not apply to any of the models
examined in this paper.}

Lanzagorta and Ross \cite{RLfixed} also considered the effect of 
various GUT theories above the scale $M_{GUT}\approx 10^{16}$ GeV.
Just as the MSSM may be analysed in the range $M_{GUT}-M_{SUSY}$
so the SUSY GUT theory was analysed in the range $M_P-M_{GUT}$,
using similar techniques. In SUSY GUT theories
containing a large number of representations,
asymptotic freedom may be lost and the gauge coupling can grow rapidly
as it approaches $M_P$. It was argued that in such a case, the fixed point
structure in the range
$M_P-M_{GUT}$ may be more important than the MSSM fixed point
\cite{RLfixed}.
It was subsequently argued that the effect of such a SUSY GUT 
would be to lead to a low energy top quark Yukawa coupling
closer to its QFP value than the MSSM expectation \cite{RLfixed}.

The fixed point nature in the SUSY GUT in the region $M_P-M_{GUT}$
is seen from the following result, obtained by analogy with earlier results,
for the top quark Yukawa coupling at the GUT scale,
\begin{equation}
R(M_{GUT})= \frac{R_{GUT}^*}
{1+ \Delta_{GUT} \left[ \frac{R_{GUT}^*}{R(M_P)} -1 \right]}
\label{GUTFPclear}
\end{equation}
where above we have replaced $t$ by its lower argument $\mu$,
in order to help keep track of which energy scale
we are referring to, and defined 
\begin{equation}
\Delta_{GUT} = \left( \frac{\alpha(M_{GUT})}{\alpha(M_P)} 
\right)^{B_{GUT}}
\label{DeltaGUT}
\end{equation}
where $B_{GUT} \equiv r/b + 1$ for the GUT theory,
and $\alpha(\mu)$ is the GUT gauge coupling at the scale $\mu$,
with $R=(Y_t/\tilde{\alpha})$.
Clearly the QFP for the GUT theory is achieved when the 
following condition is met,
\begin{equation}
R(M_P) \gg R_{GUT}^* 
\label{GUTcondition}
\end{equation}
which when satisfied leads to the approximate result
\begin{equation}
R(M_{GUT})\approx \frac{R_{GUT}^*}{1 - \Delta_{GUT}}
\equiv R_{GUT}^{QFP}
\label{GUTquasi}
\end{equation}
In the type of theories considered by Lanzagorta and Ross
\cite{RLfixed} (i.e.\ very 
non-asymptotically free GUT theories)
they find that 
\begin{equation}
R_{GUT}^*\gg R_{MSSM}^*, 
\label{GUTresults}
\end{equation}
The result in Eq.\ref{GUTresults}
implies that
the SUSY GUT is less likely to satisfy its QFP
condition in Eq.\ref{GUTcondition}.\footnote{The SUSY GUT QFP
condition is explicitly
$Y_t(M_P)\gg \tilde{\alpha}(M_P)R_{GUT}^* $
where both
$R_{GUT}^*$ and $\tilde{\alpha}(M_P)$ are
typically much larger than in the MSSM\@. This implies that
$Y_t(M_P)$ would have to be {\em substantially\/} larger
than its MSSM equivalent in order for the QFP to be relevant
for the SUSY GUT theory, leading to the danger of perturbation
theory breakdown for the top Yukawa coupling.}
For SUSY GUTs with many added vector families, 
$\Delta_{GUT}\ll \Delta_{MSSM}$
implies that the GUT FP is realised more accurately
at the GUT scale. The important question, however, is the effect
of the combination of these two results on the low energy top
quark Yukawa coupling; the effect is to drive it
more closely to its MSSM QFP value as seen below.

In order to investigate the effect of the 
SUSY GUT theory on the low energy top Yukawa coupling,
Lanzagorta and Ross first re-wrote Eq.\ref{GUTFPclear} as
\begin{equation}
x'= x\Delta_{GUT} + \frac{R_{MSSM}^*}{R_{GUT}^*}(1-\Delta_{GUT})
\label{xprime}
\end{equation}
where 
\begin{equation}
x'\equiv \frac{R_{MSSM}^*}{R(M_{GUT})},\ \ 
x\equiv \frac{R_{MSSM}^*}{R(M_P)}
\label{xdefn}
\end{equation}
The quantity $x$ should not be confused with the quantity
which gives the condition for the GUT QFP in Eq.\ref{GUTcondition}
which is
\begin{equation}
y \equiv \frac{R_{GUT}^*}{R(M_P)}
\label{y}
\end{equation}
where $y\ll 1$ is the GUT QFP condition.
The quantity $x'$ is identified as
the ratio $\frac{R^*}{R(0)}$ in Eq.\ref{FPclear},
which may consequently be written as,
\begin{equation}
R(M_{SUSY})= \frac{R_{MSSM}^*}
{1+ \Delta_{MSSM} \left[ x' -1 \right]}
\label{MSSMFPclear}
\end{equation}
The combination of Eqs.\ref{xprime} and~\ref{MSSMFPclear}
give us all the information we need to decide the fate of the
low energy top quark Yukawa coupling.
The condition for the MSSM QFP is clearly just $x'\ll 1$,
where $x'$ is given in Eq.\ref{xprime}.
According to Eq.\ref{GUTresults} we have from Eq.\ref{xprime}
\begin{equation}
x'\approx x\Delta_{GUT} 
\label{xprimeapprox}
\end{equation}
Since $\Delta_{GUT}<1$ Eq.\ref{xprimeapprox} shows that 
a given value of $x$ implies a smaller value of $x'$.
Thus the effect of such SUSY GUTs is
to make it more likely that the low energy top Yukawa coupling
is at its MSSM QFP, as claimed \cite{RLfixed}.

We now turn to the question of the infra-red
nature of  MSSM+X models. 
\footnote{In some superstring models, 
the top quark Yukawa coupling is predicted at the string scale.
For example ref.~\protect\cite{Farmt} discusses such a mechanism,
including the effects of intermediate matter.
In order to remain as model independent as possible, 
in our present analysis we shall instead 
regard $h_t$ to be unconstrained at $M_X$.} 
This approach to MSSM+X theories turns out to have many similarities
to the case of SUSY GUTs considered above; for example 
in MSSM+X theories with a large number of exotic colour triplets,
asymptotic freedom of QCD is lost above the intermediate scale.
It would therefore be expected that in such MSSM+X models,
the low energy top quark Yukawa coupling is more likely
to be at its MSSM QFP, as in SUSY GUTs, and 
we find that this is indeed the case.
The analytic results for the MSSM+X models may be more or less
taken over immediately from the SUSY GUT results given above,
by making the following obvious replacements in
Eqs.\ref{GUTFPclear}-\ref{xprimeapprox}:
\begin{equation}
M_{P}\rightarrow M_X, \ \ M_{GUT}\rightarrow M_I, \ \
R_{GUT}\rightarrow R_X, \ \ \Delta_{GUT}\rightarrow \Delta_X, \ \ 
\alpha \rightarrow \alpha_3
\label{replacements}
\end{equation}
Note that in the present case there is no fixed scale
which separates the MSSM from the MSSM+X theory, since 
$M_{GUT}$ has been replaced by the
intermediate scale $M_I$ which can range over several orders of magnitude.
This implies that $\Delta_{MSSM}$ is no longer a fixed quantity,
since it is given by,
\begin{equation}
\Delta'_{MSSM} = \left( \frac{\alpha(M_{SUSY})}{\alpha(M_I)} 
\right)^{B_{3}}
\label{DeltaMSSM}
\end{equation}
and consequently Eq.\ref{MSSMFPclear} becomes
\begin{equation}
R(M_{SUSY})= \frac{R_{MSSM}^*}
{1+ \Delta'_{MSSM} \left[ x' -1 \right]}
\label{MSSMFPclear2}
\end{equation}
where $x'$ given by
\begin{equation}
x'= x\Delta_X + \frac{R_{MSSM}^*}{R_X^*}(1-\Delta_X)
\label{xprimeX}
\end{equation}
with,
\begin{equation}
x=\frac{R_{MSSM}^*}{R(M_X)},\ 
\Delta_X=\left( \frac{\alpha(M_I)}{\alpha(M_X)}\right)^{B_{3X}} 
\end{equation}
The relevant fixed point quantities above are shown in
Table~\ref{tab:modelsfixed}. Note that $\Delta_X > \Delta'_{MSSM}$
(except in model K) where the values are comparable
to $\Delta_{GUT}$ given for various models with no extra families in
ref.\cite{RLfixed}. This is not surprising 
or even significant since $\Delta'_{MSSM}$
is calculated using a much larger ratio of scales than 
$\Delta_X$. The fact that 
$\Delta_X \ll 1$ is the important fact, and also that
$\frac{R_{MSSM}^*}{R_X^*} <1$, which implies
that $x'$ in Eq.\ref{xprimeX} is likely to be small.

\begin{table}
\begin{center}
\begin{tabular}{|c|ccccc|} \hline
Model& $R_X^*$ & $B_{3X}$&
$\frac{R_{MSSM}^*}{R_X^*}$ & $\Delta'_{MSSM}$ & $\Delta_X$\\ \hline
A &5/9&-5/3&7/10&0.56 &0.80 \\
B &13/18&-13/3&7/13&0.52 &0.87 \\
C &13/18&-13/3&7/13&0.56 &0.73 \\
D &19/18&19/3&7/19&0.57 &0.60 \\
E &8/9&*&7/16&0.51 &0.89 \\
F &8/9&*&7/16&0.52 &0.78 \\
G &11/9&11/3&7/22&0.52 &0.78 \\
H &19/18&19/3&7/19&0.50&0.90\\
I &11/9&11/3&7/22&0.57&0.54\\
J &8/9&*&7/16&0.56 &0.67 \\
K &14/9&7/3&1/2&0.57&0.39 \\
L &19/18&19/3&7/19& 0.51&0.87 \\
M &25/18&25/9&7/25&0.51 &0.83 \\
N &11/9&11/3&7/22&0.50 &0.91 \\ \hline
X &43/18&43/27&7/43&0.50 &0.78 \\ \hline
\end{tabular}
\end{center}
\caption{Fixed point properties as discussed in the text of the 
MSSM+X models defined in Table~\protect\ref{tab:models}.
$\Delta_{MSSM}'$ and $\Delta_X$ are relevant for
$\alpha_S(M_Z)=0.117$, $M_{SUSY}=500$ GeV and $m_t^{phys}=174$ GeV.}
\label{tab:modelsfixed}
\end{table}

In the case of SUSY GUTs, small values of $x'$ imply that the
MSSM QFP will be realised. Here we cannot exactly make this statement
because the MSSM is now effective below the scale $M_I$, so the
MSSM QFP here is not the same as the usual one.
In order to overcome this difficulty we 
combine Eqs.\ref{MSSMFPclear2} and~\ref{xprimeX}
into a single equation which yields the
low energy top quark Yukawa coupling directly from the string scale
boundary conditions,
\begin{equation}
R(M_{SUSY})= \frac{R_{MSSM}^*}
{1+ \Delta'_{MSSM} \left[                        
\left(x\Delta_X + \frac{R_{MSSM}^*}{R_X^*}(1-\Delta_X)\right)
-1 \right]}
\label{complicated}
\end{equation}
It is clear from Eq.\ref{complicated} that a low energy QFP 
will be achieved when the following condition is met:
\begin{equation}
x\Delta_{X} \ll 1
\label{xX}
\end{equation}
which should be compared to the MSSM QFP condition
$x'\ll 1$. Since in general $\Delta_X <1$, Eq.\ref{xX} shows that
in MSSM+X models the QFP condition is more easily achieved
than in the MSSM\@. The effect is greater for the
MSSM+X models with the smaller values of $\Delta_X$
in Table~\ref{tab:modelsfixed}. 

When the condition in Eq.\ref{xX} is satisfied, the low energy
top Yukawa coupling is given approximately independently
of its string scale input value. In other words there is a QFP
given by
\begin{equation}
R(M_{SUSY})\approx \frac{R_{MSSM}^*}
{1-  \Delta}
\label{quasiX}
\end{equation}
where
\begin{equation}
\Delta = \Delta'_{MSSM} \left[1-                        
\frac{R_{MSSM}^*}{R_X^*}(1-\Delta_X) \right]
\label{D}
\end{equation}
where we have written Eq.\ref{complicated}
in the form of Eq.\ref{quasi}, and have made the approximation 
in Eq.\ref{xX}.\footnote{For the case where the intermediate effective
theory has a zero QCD beta function, a similar expression is found
although we do not go 
into detail here.}

\begin{table}
\begin{center}
\begin{tabular}{|c|c|c|} \hline
$\alpha_3(M_Z)$&$M_{SUSY}$/GeV&$\Delta$ \\ \hline
0.112 & 200 & 0.46 \\
0.112 & 1000 & 0.51 \\
0.122 & 200 & 0.44 \\
0.122 & 1000 & 0.50 \\ \hline
\end{tabular}
\end{center}
\caption{The quantity $\Delta$ for the
MSSM and the models A,\ldots,N,X.}
\label{tab:delta}
\end{table}

The values of $\Delta$ in Eq.\ref{D}
were determined for each of these models
and were found to within the accuracy of our calculations to be
independent upon which particular
MSSM+X model was used.
In other words we find $\Delta \approx \Delta_{MSSM}$
for all of the MSSM+X models. This seems at first sight
to be somewhat surprising since $\Delta$ depends on
$\Delta'_{MSSM}$, $R_X^*$ and $\Delta_X$, all of which
vary from model to model. Somehow all these quantities
conspire to give $\Delta \approx \Delta_{MSSM}$.
The explanation is simply that
the lower energy dynamics (below $M_I$)
has the most important focusing effect on the large top Yukawa
coupling and all the higher energy differences become irrelevant. 
Thus the high energy structure of the 
MSSM+X models above the intermediate scale makes little
difference to the QFP prediction. Of course the high energy 
structure of the MSSM+X models is vital in determining
whether the top Yukawa coupling is in the QFP region
at all, as is clear from Eq.\ref{xX}.
Also it is clear that
the value of $\Delta_{MSSM}$ and hence $\Delta$
is sensitive to the
input parameters ($\alpha_3 (M_Z)$ and $M_{SUSY}$) as shown
in Table~\ref{tab:delta}. The dependence of $\Delta$ upon $m_t$ was
found to be negligible. 

\begin{figure}
\begin{center}
\leavevmode
\hbox{%
\epsfxsize=4.5in
\epsfysize=3in
\epsffile{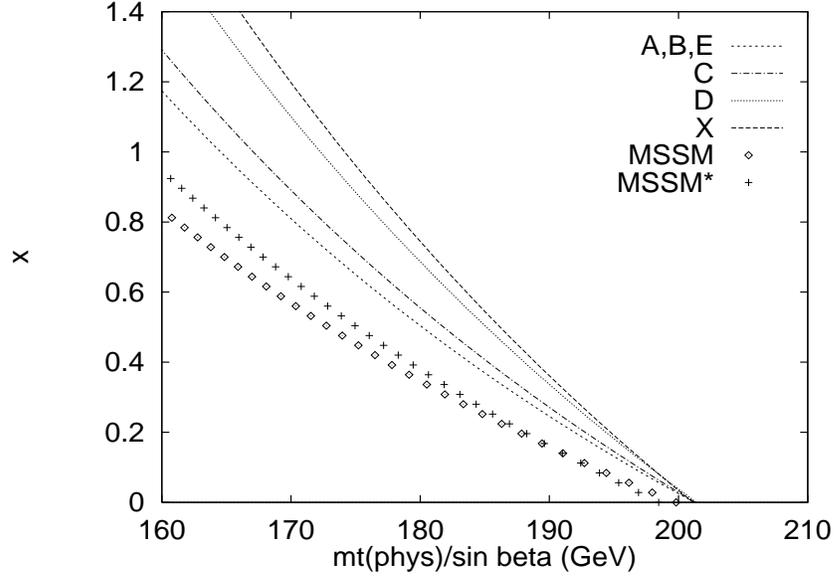}}
\end{center}
\caption{$m_t^{phys} / \sin \beta$ 
as a function of  the variable $x=\frac{R_{MSSM}^*}{R(M_X)}$
for MSSM+X models A,B,C,D,E,X and the
MSSM for
$M_{SUSY}=500$ GeV, $\alpha_3(M_Z)=0.117$. The unification scale $t=0$ for
the MSSM (MSSM*) is assumed to be $M_{GUT} =10^{16}(4.10^{17})$ GeV.}
\label{fig:lang1}
\end{figure}

\begin{figure}
\begin{center}
\leavevmode
\hbox{%
\epsfxsize=4.5in
\epsfysize=3in
\epsffile{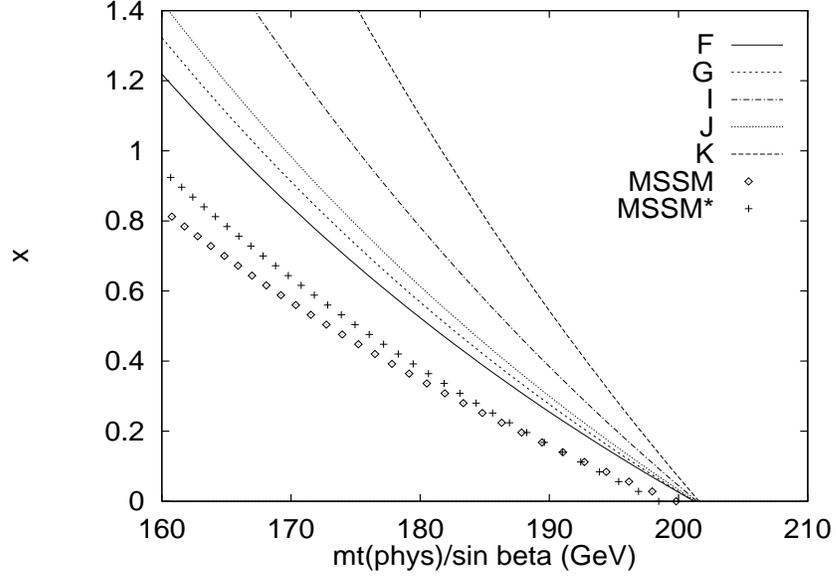}}
\end{center}
\caption{$m_t^{phys} / \sin \beta$ 
as a function of the variable $x=\frac{R_{MSSM}^*}{R(M_X)}$ for 
MSSM+X models F,G,I,J,K and the MSSM for
$M_{SUSY}=500$ GeV, $\alpha_3(M_Z)=0.117$. The unification scale $t=0$ for
the MSSM (MSSM*) is assumed to
be $M_{GUT} =10^{16}(4.10^{17})$ GeV.}
\label{fig:lang2}
\end{figure}

\begin{figure}
\begin{center}
\leavevmode
\hbox{%
\epsfxsize=4.5in
\epsfysize=3in
\epsffile{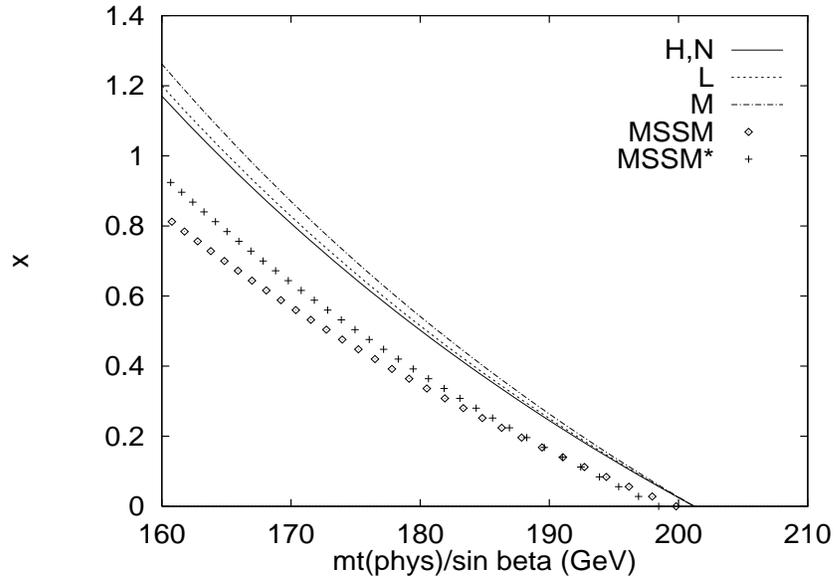}}
\end{center}
\caption{$m_t^{phys} / \sin \beta$ as a function of the 
variable $x=\frac{R_{MSSM}^*}{R(M_X)}$ for the MSSM+X
models H,L,M,N and the MSSM for
$M_{SUSY}=500$ GeV, $\alpha_3(M_Z)=0.117$. The unification scale 
corresponding to $t=0$ for
the MSSM (MSSM*) is assumed to
be $M_{GUT} =10^{16}(4.10^{17})$ GeV.}
\label{fig:lang3}
\end{figure}

The above analytic results take into account only the QCD coupling.
If electroweak corrections to Eq.\ref{FPclear} are applied
\cite{BandB}, there is
no longer an exact fixed point and the approximate quasi-fixed-point
value of
$m_t^{phys} / \sin \beta$ increases from $\sim 180$ GeV to $\sim 200$
GeV\footnote{This number is quite dependent on the input parameters.
For example, if $M_{SUSY}=2$ TeV, the quasi-fixed-point corresponds
to $m_t^{phys}/sin \beta \sim 220$
GeV.}where $m_t^{phys}$ refers to the physical (pole) mass of the
top quark. Thus these additional corrections are quite important
and must be considered.
In  Figs.\ref{fig:lang1},\ref{fig:lang2},\ref{fig:lang3} we display
the full numerical predictions for the MSSM+X and MSSM models,
obtained by numerically integrating the RG equations including
all the Higgs and electroweak couplings in addition to the
QCD coupling\footnote{For $M_{SUSY}=1$ TeV, $\alpha(M_X)=1/24$,
$M_X=1.2\times10^{16}$ GeV, the MSSM curve agrees with the plot in
ref.\cite{RLfixed}}. 
The MSSM* curve corresponds to the MSSM particle content and gauge
group up to $\mu=4\times10^{17}$ GeV, and is intended to show the
added focusing effect of increasing the range of $\mu$ by a factor of
20 compared to the MSSM\@.
The top quark mass (scaled by $\sin \beta$) is
plotted as a function of the input variable 
$x=\frac{R_{MSSM}^*}{R(M_X)}$.
Since $x$ is proportional to 
$1/Y_t(M_X)$, the zero
intercept on the horizontal axis corresponds to the quasi-fixed-points
of the models.  Note that the scale $M_X$
at which the input couplings 
$\frac{1}{R(M_X)}=\frac{\tilde{\alpha}_3(0)}{Y_t(0)}$
are defined differs from curve to curve.
The MSSM (MSSM*)has its input couplings 
defined at $10^{16}$GeV ($4\times 10^{17}$ GeV),
while the other models have $M_X$ in the range $3.5-5.5\times 10^{17}$
GeV, as shown in Fig.\ref{fig:MX}. Varying
$M_{SUSY}=200-1000$ GeV and $\alpha_3(M_Z)=0.112-0.122$ produces a maximum
(but significant) 5\% error in $m_t^{phys} / \sin \beta$.

The models in
Figs.\ref{fig:lang1},\ref{fig:lang2},\ref{fig:lang3}
corresponding to the steepest graphs
correspond to the MSSM QFP prediction of the top
quark Yukawa coupling being more likely to be realised (a vertical line
would predict the top quark mass independently of the input
Yukawa coupling.) These results may be compared to the MSSM results
which are also plotted, where in this case we have assumed 
the high energy scale to be $M_{GUT}$ so that $x=x'$ in this case.
All the MSSM+X models are steeper than the MSSM line,
indicating the increased likelihood that the QFP is realised.
The amount of the effect which is due to the extra factor
of $\sim 20$ in the range of running is illustrated by the MSSM* curve
where a higher energy scale comparable to the string scale is assumed
to be $M_{GUT}$. 
The graphs with the highest number of SU(2)$_L$ doublets (i.e.\ high
$p$)
are the steepest. 
This is in part due, however,  to the fact that
many of these models
have higher $\alpha_3(0)=\alpha_X=\alpha(M_X)$,
as is clear from Fig.\ref{fig:aGUT}. 
These plots make the focusing effect of the
fixed point clear: 
for example, model K predicts $m_t^{phys}/\sin \beta > 185$
GeV for $x<1$. These numerical results support the earlier analytical 
expectations that the smaller the value
of $\Delta_X$, the closer a particular model is likely to be
to the QFP. 

\section{Origin of Yukawa Matrices with Texture Zeroes in MSSM+X Models}

So far we have been concerned with the issues of gauge coupling
unification, and the determination of the top quark Yukawa coupling
via the infra-red fixed point structure of the MSSM+X models.
We have seen that the physical top quark mass is determined
to some extent by the infra-red fixed point of the theory,
and so the next obvious question is to what extent the 
remainder of the Yukawa matrices may be determined. We leave the
highly model-dependent possibility that the lighter fermion masses are
generated at the level of string theory alone and instead concentrate
on a possible mechanism at the effective field theory level.

Some time ago, Ibanez and Ross \cite{IR} showed how the
introduction of a gauged $U(1)_X$ family symmetry could be used 
to provide an explanation of successful quark and lepton 
Yukawa textures within the framework of the MSSM\@.
The idea is that the $U(1)_X$ family
symmetry only allows the third family to receive a renormalisable
Yukawa coupling
but when the family symmetry is broken at a scale not far below the
string scale other families receive suppressed effective Yukawa
couplings.
The suppression factors are essentially powers of 
the VEVs of $\theta$ fields which are MSSM singlets but carry
$U(1)_X$ charges and are
responsible for breaking the family
symmetry. These Yukawa couplings are scaled by heavier mass scales $M$
identified as the 
masses of new heavy vector representations which also carry
$U(1)_X$ charges. For example, one may envisage a series of 
heavy Higgs doublets of mass $M$
with differing $U(1)_X$ charges
which couple to the lighter families via sizable Yukawa
couplings which respect the family symmetry. 
The heavy Higgs doublets also couple to the MSSM Higgs doublets
via $\theta$ fields and this results in suppressed effective 
Yukawa couplings when the family symmetry is broken.
For more details of this mechanism see ref.\cite{IR}.

Recently Ross \cite{GG} has combined the idea of a gauged
$U(1)_X$ family symmetry with the previous discussion
of infra-red fixed points. The idea behind this
approach is that since there are no small Yukawa couplings
one may hope to determine all the Yukawa couplings
by the use of infra-red fixed points along similar lines
to the top quark Yukawa coupling determination. 
An explicit model was discussed in detail \cite{GG}.
The explicit model was based on the MSSM gauge group
persisting right up to the string scale.
The question of gauge coupling unification
was addressed \cite{GG}
by adding complete $SU(5)$ vector representations
to the MSSM theory with masses just below the unification scale.
These have no relative effect on the running of the three gauge couplings
to one loop order, however at two loop order 
it was claimed that the unification
scale is raised. By adding a sufficiently large number of such
states it was hoped that the unification scale could be postponed
to the string scale by a combination of two loop gauge running
and threshold effects, although this mechanism was not studied
in detail in ref.\cite{GG}. This mechanism is obviously quite different
to the one loop approach to gauge unification within the MSSM+X models
considered here, and it is clearly of interest to see if the
$U(1)_X$ family symmetry approach can be accommodated within
this class models.

In order to obtain the desired Yukawa textures it is necessary
to add several heavy Higgs doublets in vector representations
and with various $U(1)_X$ charges in addition to the two Higgs
doublets of the MSSM \cite{IR,GG}. In ref.\cite{GG} each of these
Higgs representations is accompanied by a colour triplet in order
to make up a complete $SU(5)$ representation, where such triplets
are forbidden from mixing with quarks by R-symmetry.
From our point of view, in MSSM+X models there is necessarily
additional matter at the intermediate scale $M_I$ which has
to be present in order to satisfy the condition of one loop
gauge coupling unification which we have imposed on the models.
Many of the models involve additional doublets which may be
identified with Higgs doublets if they have suitable hypercharges,
and so it is natural for us to put this extra matter to work for
us in providing the Yukawa structures. 
In principle there is no restriction on the magnitude
of the intermediate scale $M_I$,
corresponding to the masses of the extra Higgs doublets,
since it may be assumed
that the $U(1)_X$ family symmetry is broken slightly below
this scale yielding phenomenologically acceptable 
suppression ratios in the effective Yukawa couplings of $<\theta>/M_I
\sim 0.2$.
However there is a technical restriction that 
the $U(1)_X$ family symmetry should not be broken
more than a couple of orders of magnitude
below the string scale since its anomaly freedom
relies on the Green-Schwarz mechanism \cite{IR}. In fact, the
Green-Schwarz mechanism requires
\begin{equation}
\frac{\langle \theta \rangle}{M_X} \sim O(\frac{1}{\sqrt{192} \pi}),
\end{equation}
and since $\langle \theta \rangle / M_I \sim 0.2$, we must have
$M_I/M_X \sim O(1/8)$ for the mechanism to work.

\begin{figure}
\begin{center}
\leavevmode
\hbox{%
\epsfxsize=4.5in
\epsfysize=3in
\epsffile{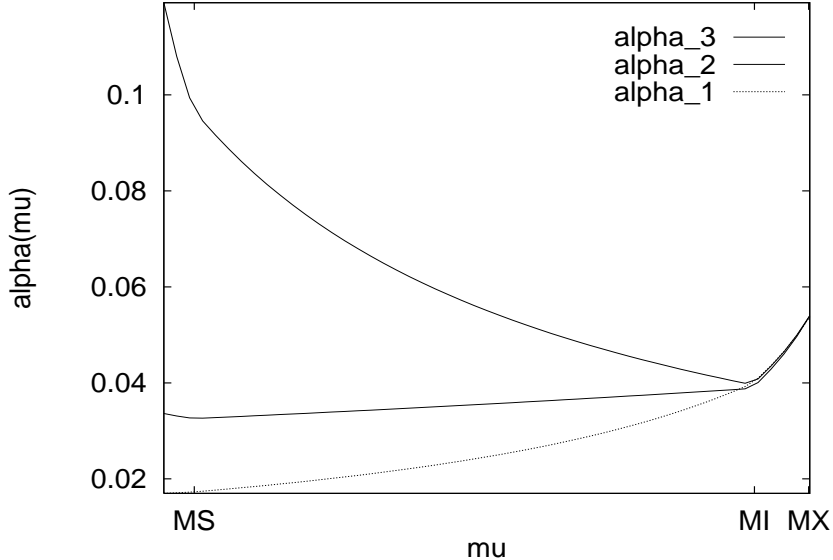}}
\end{center}
\caption{The running of the three Standard Model gauge couplings in
model X between $\mu=M_Z$ and $\mu=M_X$ for $\alpha_S(M_Z)=0.1192,
M_{SUSY}=500$ GeV, $m_t^{phys}=174$ GeV.
The normalisation $k_1=5/3$ for the hypercharge gauge coupling has
been used.}
\label{malcolmX}
\end{figure}
Let us consider as an example the model discussed
in ref.\cite{GG} in which the Higgs doublets
of the MSSM may be written as $H_1^{(0)},H_2^{(0)}$,
and the extra Higgs doublets may be written as:\footnote{In fact this
model must
necessarily involve yet more Higgs doublets 
in order to achieve Cabibbo mixing.
However for illustrative purposes we shall only consider
those Higgs doublets listed in ref.\cite{GG}.}
\begin{equation}
H_{1,2}^{-1}, \bar{H}_{1,2}^{1},
H_{1,2}^{-2}, \bar{H}_{1,2}^{2},
H_{1,2}^{3}, \bar{H}_{1,2}^{-3},
H_{1,2}^{4}, \bar{H}_{1,2}^{-4},
H_{1,2}^{8}, \bar{H}_{1,2}^{-8}, 
\end{equation}
where the $U(1)_X$ charges are given in parentheses,
and $H_{1,2}^{(x)}$ have hypercharges $Y/2=-1/2,1/2$, 
and $\bar{H}_{1,2}^{(-x)}$ have hypercharges $Y/2=1/2,-1/2$, 
respectively. The idea is that the Higgs $H_{1,2}^{(x)}$
have direct couplings to the lighter families and
mix with the MSSM Higgs $H_{1,2}^{(0)}$ via singlet 
$\theta$ fields,
thereby generating hierarchical Yukawa structures.
If the extra Higgs doublets are interpreted as intermediate
scale matter then this corresponds to $b=10$, where $b$ labels
the number of additional vector (1,2) representations.
This model may be embedded in model X which was considered previously,
and chosen with this discussion in mind.
In Model X, $M_I$ is not too far below $M_X$; choosing for example 
$\alpha_S(M_Z)=0.1192$, we have $M_I / M_X \sim 1/25$. If
we assume that 
all the other additional superfields have zero hypercharge assignment,
then we determine 
$k_1=5/3$, just right for the Green-Schwarz mechanism to work. 
These hypercharge assignments have the added advantage that they
automatically ban any mass mixing terms (above $\mu \sim M_W$) between
the ordinary quark fields and the heavy (3,1) or (3,2) fields.
Fig.\ref{malcolmX} displays the running of the three gauge couplings for
$\alpha_S(M_Z)=0.1192$.  It turns out that this model has $M_I =
1.7\times10^{16}$ GeV$\sim M_{GUT}$ and the intermediate matter
performs the job of making the couplings run with similar slope above
$M_{GUT}$.

Whereas the conditions upon $M_I/M_X$ and $k_1$ implied by the
Green-Schwarz mechanism are non-trivial to
solve in the context of the gauge unified MSSM+X models and in general
are only satisfied for some subspace of the
phenomenologically allowed values of $\alpha_S(M_Z)=0.112-0.122$ and
$M_{SUSY}$, model X is only one example of a class of possible models. For
example, by adding more doublets it is possible to increase $M_I/M_X$
in order to reach $1/8$. It may however, be possible to construct
\cite{workinprogress}
models with less particle content than model X in which the matter is
at slightly
different scales, or in which the $\theta$ field is not added
vectorially \cite{anomu1}, in order to circumvent the possible problem
of D-flatness
\cite{IR}. Here we are only concerned with presenting a model of
fermion masses that fits in with
string-scale gauge unification. 

\section{Conclusion}
We have taken the idea of intermediate scale matter to explain stringy
gauge unification seriously within the context of Kac-Moody level 1
superstrings. To make our calculation not depend upon the precise
string model chosen, we have made crude, simplifying approximations.
A strong approximation is the neglect of heavy or string threshold
effects around $M_X$. Another source of these uncertainties could come
from the assumption that the superpartner masses are all degenerate at
$M_{SUSY}$. It is well known that a significant relaxation of this
assumption may change the constraints of gauge
unification\footnote{And is likely to give flavour changing neutral
current effects excluded by experiment.}. One reason that we do not
worry too much about these possible effects is that a previous
analysis \cite{Dinesetal} showed that not only can these effects alone not
explain the discrepancy in unification of the gauge couplings at the
string scale, they sometimes tend to make the problem worse. Another
approximation we have made \cite{smoothies} is that of the step
function for particle thresholds. We do not expect this uncertainty to
be significant in a one loop calculation.

One important question which we have hitherto left unanswered is: where
does the extra matter come from and why does it have the mass $M_I\ll
M_X$?
There are several possible solutions to this question: it is possible
that it is connected to some sort of hidden sector dynamics
\cite{GUST}, possibly
related to supersymmetry breaking. Another possibility is of some
non-renormalisable operators coming straight from the string
\cite{Dinesetal}. Models
with high additional numbers of right handed quark fields have $M_I$
about an order of magnitude below $M_X$ and so the problem of how the
$M_I-M_X$ hierarchy arises in these models may not be relevant.

We have systematically analysed what constraints there
exist on the extra matter when it all has roughly an equivalent mass
for up to 5 vector representations additional to the MSSM\@.
Predictions for the scales $M_I, M_X$ are given by the string gauge
unification conditions. It is also found that the number of extra
right handed quark type fields must exceed the number of the extra
left handed quark and lepton (or Higgs)
supermultiplets if the gauge couplings
are to unify at the string scale. We emphasise again that, unlike
the analysis in ref.\cite{MR}, we have not imposed the GUT normalisation
value of $k_1$ on the models,
so the identification of the extra matter
with exotic quarks and leptons or Higgs doublets is 
for descriptive purposes only since the hypercharge
assignments are arbitrary.
In fact we have obtained upper bounds
on $k_1$ for each of the models under consideration.
The theoretical lower bound of $k_1>1$ has also been used
to place a restriction on the sum of the squared hypercharges
of the additional matter for each of the models.

A large part of this paper has been concerned with
the top quark Yukawa coupling fixed points in MSSM+X
models. The effect of the additional matter
above the intermediate scale is seen to make
the MSSM QFP low energy prediction of the top 
quark mass more likely than in the MSSM\@,
with the result that the physical top quark
tends to be heavier.
In this respect the MSSM+X models
behave rather similarly to the SUSY GUT theories
which contain a large number of representations \cite{RLfixed}.
We studied this effect both analytically,
using the simple approximation of retaining only
the QCD gauge coupling constant, and numerically
keeping all three gauge couplings.
The full numerical solutions
for the top quark mass in the MSSM+X models 
are given in Figs.\ref{fig:lang1},\ref{fig:lang2},\ref{fig:lang3}.
One way of summarising our results is to say that, once the MSSM
is correctly adjusted in order to give string unification, the
top quark mass is more likely to be determined by its MSSM QFP
than in the standard MSSM\@.  Of course the QFP
prediction itself is the same in both the MSSM and the MSSM+X
models; it is just that in the MSSM+X models the QFP prediction
is more likely to realised.

The problem of the origin of the lighter fermion masses was also discussed
briefly in the context of Abelian family gauge symmetries. An example
is found, which we referred to as Model X,
which is phenomenologically acceptable as a candidate for
this scenario, having the properties that $k_1=5/3$
and an intermediate scale not too far below the string scale
as shown in Fig.\ref{malcolmX}. In Model X, ten of the
extra vector representations are identified 
as extra Higgs doublets and are assigned the appropriate hypercharges.
Following the scenario of refs.\cite{IR,GG}, the gauged $U(1)_X$
family symmetry is assumed to be broken, leading to mixing of the
standard and extra Higgs doublets, and resulting in small 
effective Yukawa couplings and approximate texture zeroes,
once suitable family charges are assumed. 
There are undoubtedly more examples of a similar
nature in addition to Model X \cite{workinprogress}.
Needless to say, in common with the other MSSM+X models,
Model X also favours the MSSM QFP prediction
of the top quark mass. 

To conclude, we find the fusion of the MSSM+X approach
to gauge coupling unification and the $U(1)_X$ gauged family
symmetry and infra-red fixed point
approach to fermion masses to be a very promising and exciting
area which deserves further study.


\begin{thebibliography}{99}
\bibitem{GUTun}
U.Amaldi et al., Phys. Rev. {\bf D36} (1987) 1385;\\
P.Langacker and M.Luo, Phys. Rev. {\bf D44} (1991) 817;\\
J.Ellis,S.Kelley and D.V.Nanopoulos, Phys. Lett. {\bf B249} (1990)
441;\\ Nucl. Phys. {\bf B373} (1992) 55;\\
U.Amaldi, W. de Boer and H.Fustenau, Phys. Lett. {\bf B260} (1991)
447;\\
C. Giunti, C.W. Kim and U.W. Lee, Mod. Phys. Lett. {\bf A6} (1991)
1745;\\
H.Arason et al., Phys. Rev. {\bf D46} (1992) 3945;\\
F.Anselmo, L.Cifarelli, A.Peterman and A.Zichichi, Nuovo Cimento 105A
(1992) 1179;\\
P.Langacker and N.Polonski, Phys. Rev. {\bf D47} (1993) 4028;\\
A.E.Faraggi and B.Grinstein, Nucl. Phys. {\bf B422} (1994) 3.
\bibitem{HR}
L. Hall and S. Raby, Phys. Rev. {\bf D51} (1995) 6524.
\bibitem{gaugeun}
J.Ellis, S. Kelley and D.V.Nanopoulos, Phys. Lett. {\bf B249} (1990)
241;\\
C.Bachas, C.Fabre and T.Yanagida, NSF-ITP-95-129, CPTH-S379.1095,
hep-th/951004;\\
P.H.Chankowski, Z.Pluciennik, S. Pokorski and C.E.Vayonakis, Phys.
Lett. {\bf B358} (1995) 264;\\
A. de la Macorra, Phys. Lett. {\bf B341} (1994) 31;\\
H.P.Nilles, TUM-HEP-234/96, SFB-375/28.
\bibitem{kap}
V.Kaplunovsky, Nucl. Phys. {\bf B307} (1988) 145.
\bibitem{su5xsu5}
C.Bachas and C.Fabre, hep-ph/9505318;\\ A.A.Maslikov, I.A.Naumov and
G.G.Volkov, hep-ph/9512429.
\bibitem{level2}
D. Lewellen, Nucl. Phys. {\bf B337} (1990) 61;\\
A. Font, L. Ibanez and F. Quevedo,
Nucl. Phys. {\bf B345} (1990) 389;\\
S. Chaudhuri, S-w Chung and J. Lykken, hep-ph/9405374;\\
G.Aldazabal, A.Font, L.E.Ibanez and A.M.Uranga, Nucl. Phys. {\bf B452}
(1995) 3;\\
I.A.Antionadis, J.Ellis, J.Hagelin and D.V.Nanopoulos, Phys. Lett.
{\bf B194} (1987) 231;\\ I.A.Antionadis, J.Ellis, J.Hagelin and
D.V.Nanopoulos, Phys. Lett. {\bf B231} (1989) 65;\\
I.A.Antionadis and G.K.Leontaris, Phys. Lett. {\bf B216} (1989) 33;\\
I.A.Antionadis, G.K.Leontaris and J.Rizos, Phys. Lett. {\bf B245}
(1990) 161.
\bibitem{ibanez}
L.E.Ibanez, Phys. Lett. {\bf B318} (1993) 73.
\bibitem{thresholds}
Choi and Kiwoon, Phys. Rev. {\bf D37} (1988) 1564;\\
P. Mayr, H.P.Nilles and S.Steinberger, Phys. Lett, {\bf B317} (1993)
53;\\
D.Bailin and A.Love, Phys. Lett. {\bf B292} (1992) 315;\\
E. Halyo, hep-ph/9509323; \\
V.Kaplunovsky, Nucl. Phys. {\bf B307} (1988) 145;\\
I.A.Antionadis, J.Ellis, R.Lacaze and D.V.Nanopoulos, Phys. Lett. {\bf
B268} (1991) 188;\\
J.P.Deredinger, S.Ferrara, C.Kounnas and F.Zwirner, Phys. Lett. {\bf
B271} (1991) 307;\\
J.P.Deredinger, S.Ferrara, C.Kounnas and F.Zwirner, Nucl. Phys. {\bf
B372} (1992) 145;\\
L.Ibanez, D.Lust and G.G.Ross, Phys. Lett. {\bf B272} (1991) 251;\\
L.Dixon, V.Kaplunovsky and J.Louis, Nucl. Phys. {\bf B335} (1991)
649;\\
I.A.Antionadis, E.Gava, K.S.Narain and T.Taylor, Nucl. Phys. {\bf
B407} (1993) 706;\\
E.Kiritsis and C.Kounnass, Nucl. Phys. {\bf B442} (1995) 472;\\
V.Kaplunovsky and J.Louis, Nucl. Phys. {\bf B444} (1995) 191;\\
P.M.Petropoulos and J.Rizos, hep-th/9601037.
\bibitem{Dinesetal}
K.R.Dienes and A.E.Faraggi, Phys. Rev. Lett. {\bf 75} (1995) 2646;\\
K.R.Dienes and A.E.Faraggi, Nucl. Phys. {\bf B457} (1995) 409.
\bibitem{ellisetc}
J.Ellis, S.Kelley and D.V.Nanopoulos, Phys. Lett. {\bf B260} (1991)
131;\\
I. Antoniadis, J. Ellis, S. Kelley and D.V. Nanopoulos,
Phys. Lett. {\bf B272} (1991) 31;\\
D.Bailin and A.Love, Phys. Lett. {\bf B280} (1992) 26;\\
D.Bailin and A.Love, Mod. Phys. Lett. {\bf A7} (1992) 1485;\\
J.Lopez and D.V.Nanopoulos, DOE-ER-40717-20, CTP-TAMU-45-95,
ACT-16-95, hep-ph/9511426;\\
A.E.Faraggi, Phys. Lett. {\bf B302} (1993) 202;\\
J.Lopez, D.V.Nanopoulos and K.Yuan, Nucl. Phys. {\bf B335} (1990) 347;\\
M.K.Gaillard and R.Xiu, Phys. Lett. {\bf B296} (1992) 71;\\
I.A.Antionadis and K.Benakli, Phys. Lett. {\bf B295} (1992) 219;\\
R.Xiu, Phys. Rev. {\bf D49} (1994) 6656.
\bibitem{MR}
S.Martin and P.Ramond, Phys. Rev. {\bf D51} (1995) 6515.
\bibitem{PR}
B.Pendleton and G.Ross, Phys. Lett.{\bf B98} (1981) 291.
\bibitem{H}
C.Hill, Phys. Rev. {\bf D24} (1981) 691.
\bibitem{RLfixed}
M.Lanzagorta and G. Ross, Phys. Lett. {\bf B349} (1995) 319.
\bibitem{IR}
L.Ibanez and G.Ross, Phys. Lett. {\bf B332} (1994) 100.
\bibitem{GG}
G.G.Ross, Phys. Lett. {\bf B364} (1995) 216.
\bibitem{originmassnonren}
A.E.Faraggi, Phys. Rev. {\bf D46} (1992) 3204.
\bibitem{GUST}
J.L.Lopez and D.V.Nanopoulos, CTP-TAMU-41/95, DOE/ER/40717 19,
ACT-15/95, hep-ph/9511266;\\
G.K.Leontaris and N.D.Tracas, IOA 327/95, NTUA 53/95, hep-ph/9511280;\\
G.K.Leontaris, Phys. Lett. {\bf B281} (1992) 54;\\
R.Barbieri, G.Dvali and A.Strumia, Phys. Lett. {\bf B33} (1994) 79;\\
J.L.Lopez, D.V.Nanopoulos and A.Zichichi, CTP-TAMU-01/96,
DOE/ER/ 40717-24, ACT-01/96, hep-ph/9601261.
\bibitem{nonst}
K.Dienes, A.Faraggi and J.March-Russell, hep-th/9510223.
\bibitem{CDF}
CDF collaboration, Phys. Rev. Lett. {\bf 73}, (1994) 225.
\bibitem{Farmt}
A.E.Faraggi, hep-ph/9506388.
\bibitem{BandB}
V.Barger, M.S.Berger and P.Ohmann, Phys. Rev. {\bf D47} (1993) 1093.
\bibitem{workinprogress}
B.C.Allanach and S.F.King, work in progress.
\bibitem{anomu1}
P.Bientruy, S.Lavignac and P.Ramond, LPTHE-ORSAY 95/54,
UFIFT-HEP-96-1,hep-ph/9601243.
\bibitem{smoothies}
L.Clavelli and P.W.Coulter, UAHEP-954, hep-ph/9507261.
\end{thebibliography}
\end{document}